\documentclass[twocolumn,nofootinbib,prl,showpacs,preprintnumbers,amsmath,amssymb,longbibliography]{revtex4-2}

\usepackage{graphicx}
\usepackage{dcolumn}
\usepackage{bm}
\usepackage{hyperref}
\usepackage{miller}

\hypersetup{
	colorlinks   = true, 
	urlcolor     = blue, 
	linkcolor    = blue, 
	citecolor   = red 
}

\usepackage[usenames,dvipsnames]{color}

\usepackage[normalem]{ulem}

\newcommand{\commout}[1]{}

\usepackage[italicdiff]{physics}

\begin{document}

\def\blue{\textcolor{blue}}
\def\gr{\textcolor[rgb]{0,0.6,0}}
\def\br{\textcolor[rgb]{0.7,0.5,0}}
\def\prp{\textcolor[rgb]{0.5,0,0.5}}

\def\LDP{LD-phase}
\def\HDP{HD-phase}
\def\AB{{\it A-B}}
\def\Ef{$E_{\rm F}$}
\def\Tc{$T_{\rm C}$}
\def\kpara{{\bf k}$_\parallel$}
\def\kparax{{\bf k}$_{\parallel,x}$}
\def\kparay{{\bf k}$_{\parallel,y}$}
\def\kperp{{\bf k}$_\perp$}
\def\dirGX{$\overline{\rm \Gamma}-\overline{\rm X}$}
\def\dirGY{$\overline{\rm \Gamma}-\overline{\rm Y}$}
\def\dirGK{$\overline{\rm \Gamma}-\overline{\rm K}_{Ag}$}
\def\dirGM{$\overline{\rm \Gamma}-\overline{\rm M}_{Ag}$}
\def\dirGS{$\overline{\rm \Gamma}-\overline{\rm S}$}
\def\dirGN{$\overline{\rm \Gamma}-\overline{\rm N}$}
\def\dirGNhalf{$\frac{1}{2}(\overline{\rm \Gamma}-\overline{\rm N})$}
\def\pntG{$\overline{\rm \Gamma}$}
\def\pntM{$\overline{\rm M}_{Ag}$}
\def\pntMprime{$\overline{\rm M'}$}
\def\pntK{$\overline{\rm K}_{Ag}$}
\def\pntN{$\overline{\rm N}$}
\def\pntNhalf{ $\overline{\rm N}/2$ }
\def\invA{\AA$^{-1}$}
\def\DCgamma{${\rm DC}_{\overline{\Gamma}}$}
\def\DCNhalf{${\rm DC}_{\overline{\rm N}/{\rm 2}}$}

\def\root33{$\sqrt{3}\times\sqrt{3}$ {\it R}30$^\circ$}
\def\RT3{$\sqrt{3}$}

\renewcommand{\andname}{\ignorespaces}

\title{One-dimensional electronic structure of phosphorene chains}

\author{Maxim Krivenkov$^{1,\dagger}$}\email[Corresponding author: ]{maxim.krivenkov@helmholtz-berlin.de}
\author{Maryam Sajedi$^{1,\dagger}$}
\author{Dmitry Marchenko$^1$}
\author{Evangelos Golias$^2$}
\author{Matthias Muntwiler$^3$}
\author{Oliver Rader$^{1}$ and Andrei Varykhalov$^{1}$}

\affiliation{$^1$ Helmholtz-Zentrum Berlin f\"ur Materialien und Energie,Elektronenspeicherring BESSY II, Albert-Einstein-Str. 15, 12489 Berlin, Germany}
\affiliation{$^2$ MAX IV Laboratory, Lund University, Fotongatan 2, 22484, Lund, Sweden}
\affiliation{$^3$ Paul Scherrer Institute, 5232 Villigen, Switzerland}
\affiliation{$^\dagger$ These authors contributed equally to this work}

\begin{abstract}
Phosphorene, a 2D allotrope of phosphorus, is technologically very appealing because of its semiconducting properties and narrow band gap.
Further reduction of the phosphorene dimensionality may
spawn exotic properties of its electronic structure, including 
lateral quantum confinement and topological edge states.
Phosphorene atomic chains self-assembled on Ag(111) have recently been characterized structurally but were found by angle-resolved photoemission (ARPES) to be electronically 2D. We show that these chains, although aligned equiprobably to three \hkl<1 -1 0> directions of the Ag(111) surface, can be characterized by ARPES because the three rotational variants are separated in the angular domain.
The dispersion of the phosphorus band  measured along and perpendicular to the chains reveals pronounced electronic confinement resulting in a 1D band, 
flat and dispersionless perpendicular to the chain direction in momentum space.
Our density functional theory calculations reproduce the 1D band for the experimentally determined structure of P/Ag(111). We predict a semiconductor-to-metal phase transition upon increasing the density of the chain array so that a 2D structure would be metallic.

\end{abstract}

\maketitle

\section{Introduction}

Since the first experiments with exfoliated graphene, the field of two-dimensional (2D) quantum materials
is in the expansive spotlight of fundamental and applied research. Post-graphene 2D materials are formed not only by group IV elements, but also group V, where the latter demonstrate semiconducting properties: prominent examples are bismuthene \cite{Bismuthene_Reis2017} and recently synthesized arsenene \cite{Shah2020}. One of the most intriguing elements of this type is phosphorus and its allotropes, revealing diversity of novel electronic and
optical properties. 
There are several structural polymorphs such as red, green and blue phosphorus.
Black phosphorus (BlackP) is the most stable one, it is a layered material composed of buckled honeycomb sheets of P atoms, held together by van der Waals interaction
\cite{BlackP-Dresselhaus-2015,BlackP-Han-2014,BlackP-Xia-2019,BlackP-Golias-2016}. 
The tunability of the band gap in BlackP in wide range (0.3$-$2.0~eV), combined with high carrier mobility, makes BlackP a perspective material for electronic design \cite{Kim-Science-2015} and it has already given promising results in transistor and optoelectronic devices \cite{Li2014,Xia2014}. However, BlackP is difficult to grow epitaxially, therefore, the mechanical/liquid exfoliation techniques used to produce few- and single-layer BlackP limit the bottom-up approach that is needed for integrated circuit development.

Another 2D polymorph with structural and electronic similarities to monolayer BlackP is the blue phosphorene (BlueP). 
So far formation of BlueP-like structures has been only reported on noble metal substrates \cite{BlueP-Zhang-Nanoletters-2016,BlueP-Zhang-Small-2018,PAu100_Schaal2021,BlueP_Ag111_Yang2020,Zhang2022}. 
In contrast to the bulk and bulk exfoliated systems, epitaxially grown 2D materials cannot easily be characterized optically, in particular when grown on metals substrates. Hence, the discovery of BlueP proceeded in two steps. At first, the candidate material monolayer P/Au(111) has been investigated structurally by STM \cite{BlueP-Zhang-Small-2018,BlueP-Zhang-Nanoletters-2016}. In a second step, angle-resolved photoemission investigated the band dispersion and confirmed the band gap of $\sim$1.1~eV of BlueP \cite{BlueP-Zhuang-ACSNano-2018,BlueP-Golias-Nanoletters-2018}. 

A major further step in the design and control of electronic structure of phosphorus would be
further reduction of phosphorene dimensionality. Theory
predicts that 1D P should host a plethora of
interesting phenomena, including giant Stark effect \cite{Wu2015} and strain-induced topological phase transition \cite{Sisakht2016}. 
It is, however, difficult to produce 1D atomic chains and also nanowires from 2D materials since their basis is their weak perpendicular but strong lateral interaction. Nevertheless, nanoribbons have been produced successfully from graphene \cite{Cai2010}, phosphorene \cite{Watts2019} and TMDCs \cite{Lim2022}. 

Recently, it was shown that
phosphorus on Ag(111) can self-assemble into 1D atomic chains with an armchair structure 
\cite{BlueP-Zhang-NatComm-2021}. Band structure of an equidistant array of 
such chains 
was studied by micro-focused angle-resolved photoemission (ARPES) \cite{BlueP-Zhang-NatComm-2021}. 
It was found to be 2D and not sharp, which led to the conclusion that 
phosphorus chains are only \textit{quasi}-1D and exhibit a non-negligible mutual 
lateral interaction \cite{BlueP-Zhang-NatComm-2021}. 
It should be noted that there are not many cases also for metallic 1D systems where the electronic confinement to 1D on the surface could be proven by ARPES, most notable ones are nanowires of Au on stepped Si(111) \cite{Segovia1999,Crain2003}, In \cite{Yeom1999,Ahn2004} and Pb \cite{Tegenkamp2008,Kim2007} on Si(111), Tb on Si(110) \cite{Appelfeller2023}, Au/Ni(110) \cite{Pampuch2000} and a special case of Bi(441) where two 1D surface-localized states were observed \cite{Bianchi2015}.

Due to the small size of structural domains even applying micro-focused ARPES at the cost of lower intensity (compared to conventional ARPES) does not guarantee to capture a single local domain of the structure.
In the present work we use conventional ARPES and demonstrate that 
atomic chains of P on Ag(111) exhibit very pronounced and perfectly 1D 
electronic structure. Phosphorus bands disperse between 3 and 1.5 eV binding energy ({\it E$_{B}$})
in the direction along the chains, but remain extraordinary flat (within 20 meV)
in the direction perpendicular to the chains at any {\it E$_{B}$}. This result is
accurately confirmed by density functional theory (DFT) calculations.
Using DFT we also study the evolution of the phosphorus bands
with changing separation between P chains, and find a 
1D$\rightarrow$2D electronic structure 
transition for marginally separated chains, which renders them a 2D metal.

\begin{figure}[!th]
\centering
\includegraphics[width=\columnwidth]{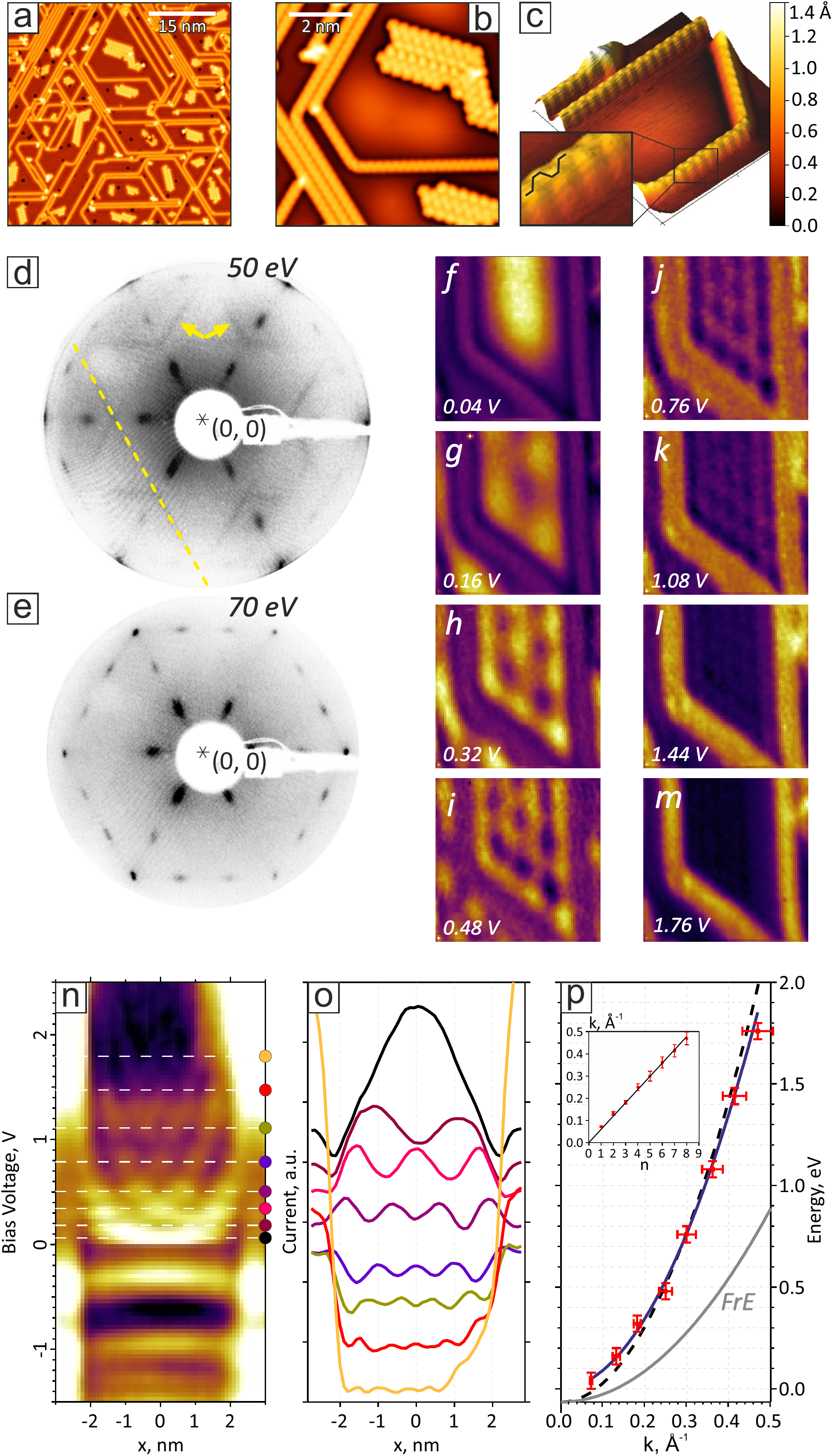}  
\caption{
(a,b) STM images of 1D P/Ag(111) with small islands of the 2D phase and (c) close-up showing armchair-type internal structure of chains (8$\times$8 nm$^2$).
(d,e) LEED
and (f--p) STM/STS characterization of P chains on Ag(111);
(f--m) $\dv{I}{V}$ maps 
showing electronic standing waves of 
Ag(111) surface state in the potential well of P chains;
(n) Differential conductance as a function of energy and position across the chains (signal is integrated along the chains), (o) its line profiles [dashed lines in (n)] and (p) extracted dispersion of the surface state (red crosses). Parabolic fit of dispersion (blue line) is compared to Ref.~\cite{AgSS_Grothe2013} (black dashed line). Free electron dispersion (FrE) with the same origin is shown in grey.
Inset shows linear $k(n)$ dependence corresponding to a potential well width of $d=5.5$ nm;
Tunneling parameters: (a) -1.0 V, 0.1 nA; (b) -0.1 V, 1.0 nA; (c) -0.1 V, 3.0 nA.
}
\end{figure}

\begin{figure}[t]
\centering
\includegraphics[width=\columnwidth]{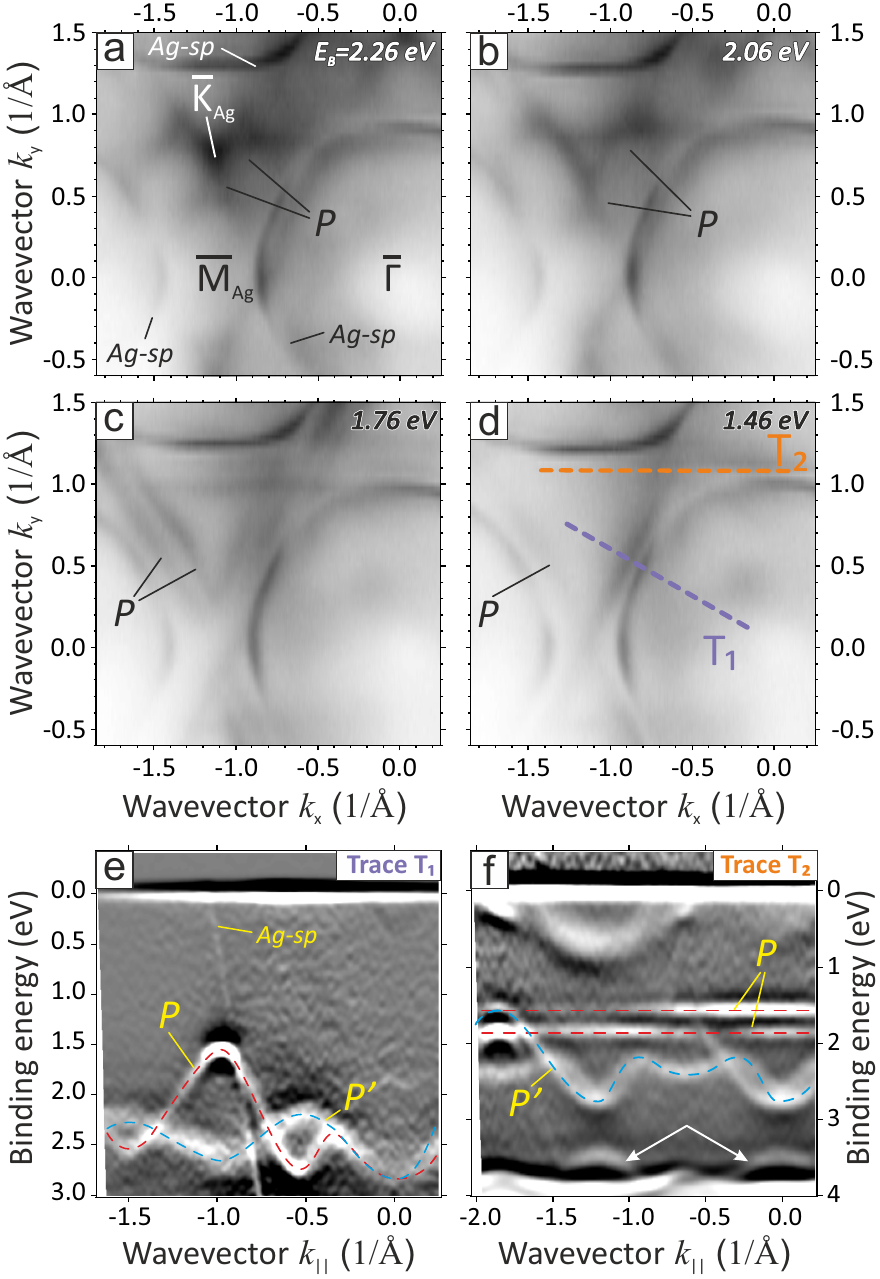} 
\caption{
(a--d) Constant energy surfaces of 1D P on Ag(111) at different binding energies {\it E$_{B}$} (h$\nu=80$ eV).
Signal from P chains is labeled as {\it P}, from Ag(111) -- as {\it Ag-sp};
(e) Dispersion of band {\it P} along the chains (h$\nu=60$ eV, $\dv[2]{I}{E}$), and (f) 
perpendicular to the chains direction and passing through the P valence band maximum (h$\nu=80$ eV, $\dv{ln(I)}{E}$). Contributions from other rotated domains are labeled {\it P'}. White arrows mark top of the next P band.
}
\end{figure}

\begin{figure}[t]
\centering
\includegraphics[width=\columnwidth]{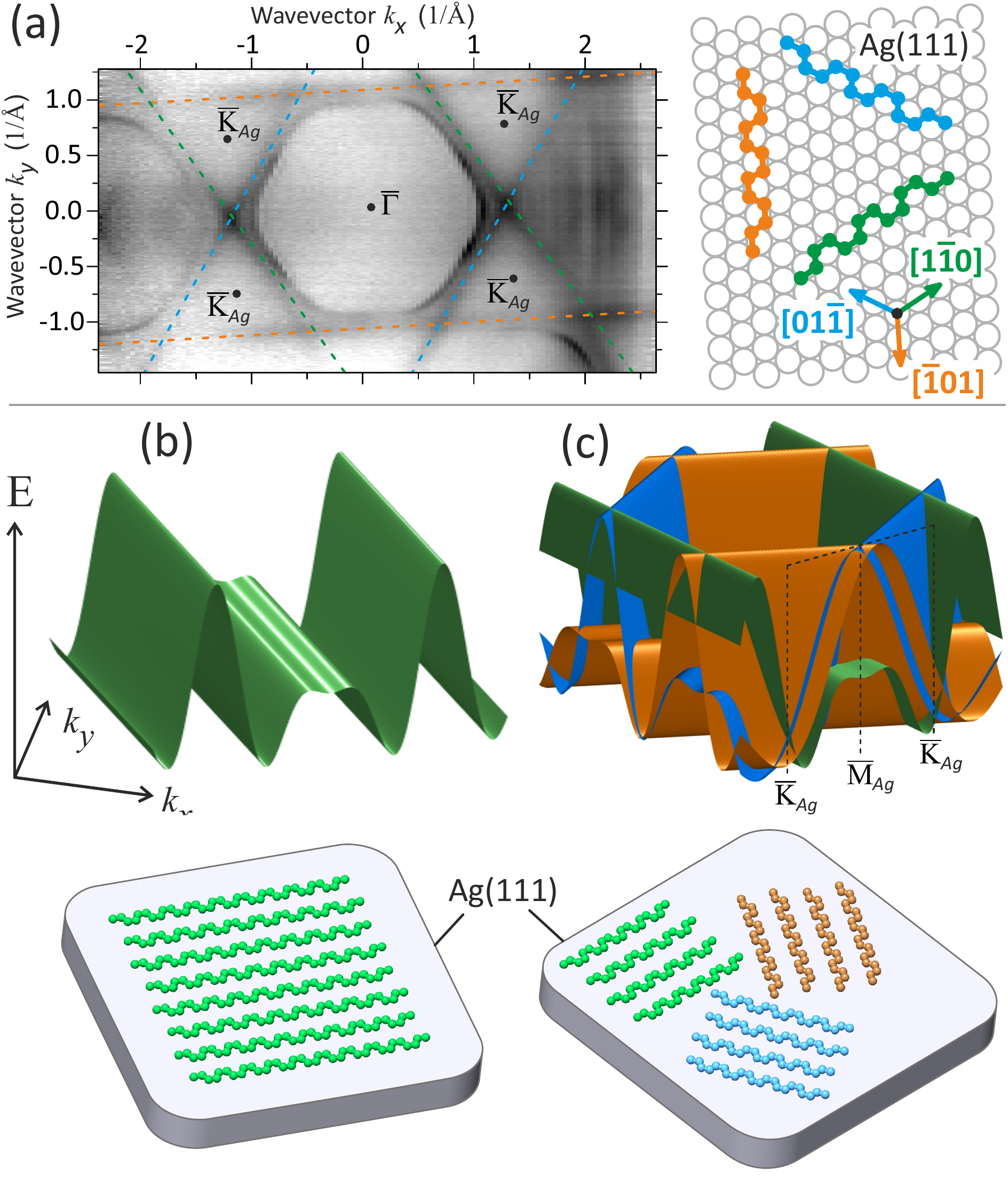} 
\caption{
(a) Plot explains constitution of the P signal in ARPES
and correspondence of straight lines in ARPES map (left) to
three orientations of chains (right). ARPES map was measured with h$\nu=100$ eV, E$_B=1.51$ eV (top of P band).
(b) Rough sketch of 1D band of P chains as referred to the orientation of the array of chains;
(c) Model of superposition of P bands from three orientational domains 
of phosphorene chains.
}
\end{figure}

\begin{figure*}[t]
\centering
\includegraphics[width=\textwidth]{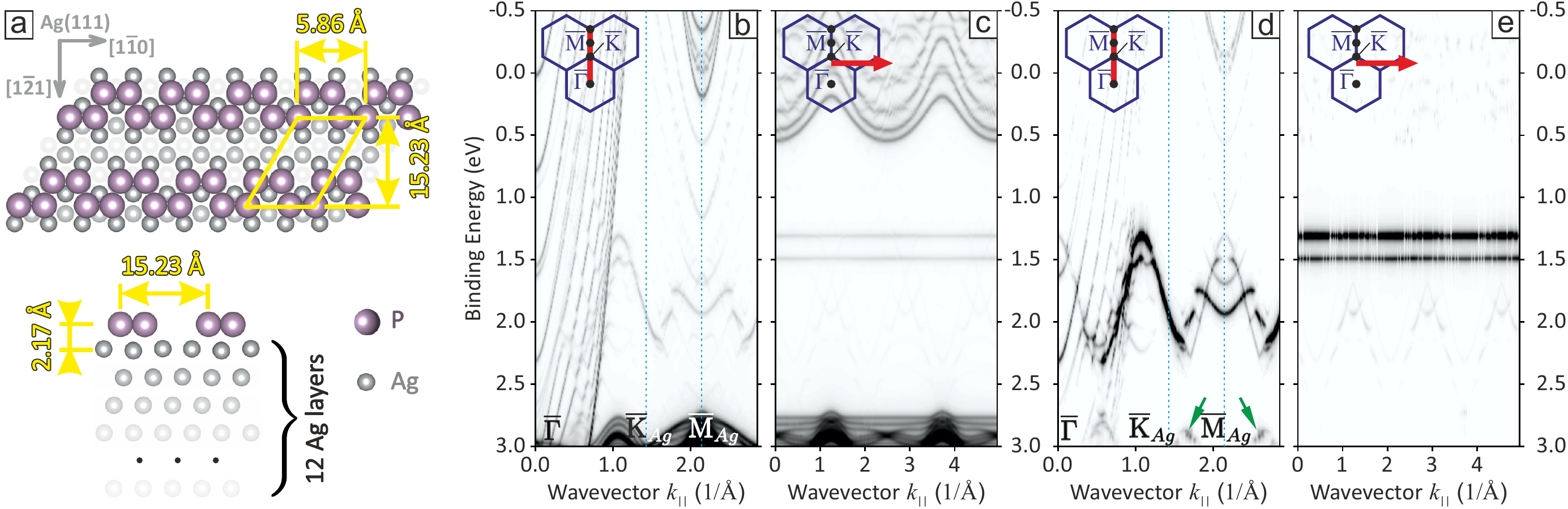}
\caption{
DFT study of P chains; 
(a) Structural model of Ag(111)/P-(2$\times$3) used in the simulation of experimentally realized chains, top-view (top) and side-view along the chains (bottom);
(b,c) Combined bands of P and Ag; (d,e) Projection of the band structure on P atoms.
The band structure was calculated along \dirGK\ of Ag(111) which is parallel to the chains (b,d) and along the perpendicular direction passing through the valence band maximum (c,e), the directions are marked in the insets by red lines. P-related bands demonstrate no dispersion perpendicular to the chains in perfect agreement with ARPES experiment. Green arrows in panel (d) mark a lower energy band of P for which a corresponding band is also found in the ARPES data. See Fig.~S2 of supporting information \cite{Supplement} for DFT plots in a wider energy range.
}
\end{figure*}

\begin{figure}[t]
\centering
\includegraphics[width=\columnwidth]{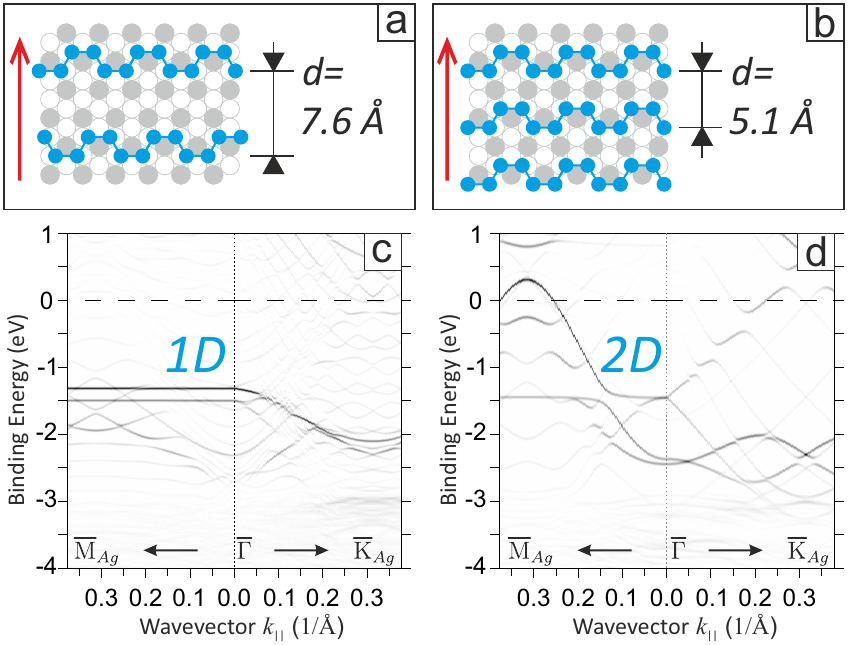} 
\caption{
1D$\rightarrow$2D electronic structure transition upon lateral compression of the chain array; (a,b) atomic model of P chains separated by $d=7.6$ \AA\ (as in our experiment) and $d=5.1$ \AA\ (artificially compressed structure); (c,d) calculated electronic band structures projected on P atoms in directions perpendicular (red arrow, \dirGK) and parallel (\dirGM) to chains.
Dispersions are displayed without unfolding.
}
\end{figure}

Information on experimental techniques and DFT calculations is provided in the supporting information \cite{Supplement}.

\section{Results and discussion}
 
Figs. 1(a,b) display characterization of P chains on Ag(111)
by scanning tunneling microscopy (STM) at moderate P coverage ($\sim$0.6 ML)
and at different length scales. P forms mainly 1D structures; however, there are some small 2D islands with lattice rotated by 30$^\circ$ relative to the chains. Fig. 1(c) shows a 3D close-up of the chains, which reveals their armchair atomic structure. From STM images one sees that
P chains are aligned along three equivalent orientations with a mutual  
angle of 120$^\circ$ between them. These orientations correspond to the 
\hkl<1 -1 0> family of directions of Ag(111). The formation of three orientational domains of P 
chains is also confirmed by low-energy electron diffraction (LEED) [Figs. 1(d,e)].
The displayed LEED patterns correspond to a sample with high P coverage, similar to the sample that was used for ARPES measurements. 
The LEED reveals a (2$\times$3) pattern which is a superposition of three equally 
intense (3$\times$1) superlattices and weak (1$\times$2) lines (marked by arrows and dashed line), corresponding to the chains in each of the three domains. The complex appearance of the LEED pattern is discussed in the supporting information \cite{Supplement}.
In this scenario, the elongation of spots can be explained by an imperfection of 
chain stacking in the directions perpendicular to the chains. 
XPS spectra of P {\it 2p} and Ag {\it 3d} core levels (Fig.~S1 of supporting information \cite{Supplement}) are in agreement (within 0.1 eV) with Ref. [\onlinecite{BlueP-Zhang-NatComm-2021}]. According to Ref. \cite{BlueP-Zhang-NatComm-2021} the fact that Ag {\it 3d} peaks show no chemical shifts as compared to bare Ag implies weak interaction between chains and the substrate and rules out P-Ag surface alloying. We observe only a small shift of Ag {\it 3d} core levels to higher {\it E$_{B}$} for chains (80 meV) and larger shift (130 meV) for 2D phase of phosphorus on Ag(111) (Fig.~S1).
The lack of Ag-P surface alloy is indirectly confirmed in scanning tunneling spectroscopy (STS) experiments, in particular 
$\dv{I}{V}$ mapping of the Ag(111) surface state (SS) scattering at P chains. 
A series of quantum interference patterns (QIP) acquired 
differentially at various binding energies (bias voltages) is displayed in Figs. 1(f--m).
The real-space periodicity of standing electron waves, converted to momentum space [Figs. 1(n--p)], allows to determine the {\it E(k)} dispersion of Ag(111) SS 
\cite{STS-QIP-SS,AgSS_Grothe2013} laterally confined between P chains. A parabolic fitting of the dispersion provides effective mass $m_{eff} = (0.46 \pm 0.03)\ m_0$ and apex energy $E_0 = (3 \pm 18)$ meV.
The dispersion was found to be similar to that of the SS at bare Ag(111) \cite{AgSS_Grothe2013}, except for 50 meV lower binding energy: $m_{eff} = (0.41 \pm 0.02)\ m_0$ and $E_0 = (-65 \pm 1)$ meV. This proves that the lateral potential barrier of P chains is so large that the particle-in-a-box model of quantization applies here.
This together with our XPS results, which exclude P-Ag surface alloying, demonstrates that there is no leaking of Ag(111) SS into the bulk at the sites of the chains.
We conclude that P chains reside on top of Ag(111) and interact with the substrate.

The photoelectrons in ARPES are collected from the whole area illuminated by the photon beam. Even with the use of micro-focused ARPES, discerning a signal from a single domain might be very challenging because of the small average size of uniaxially aligned domains.
In the present experiment beam size is on the scale of $\sim$100~$\mu$m. This means that signals from all three orientational domains of P chains are accumulated and
superimposed in the ARPES dispersions. 
For ARPES investigation of 1D structures
it is optimal to have them identical and aligned along a single
direction in order to be able to clearly separate electronic bands dispersing along the chain and non-dispersing bands of electrons confined in the perpendicular direction.

However, we show below that in the current case conventional ARPES is well applicable for mapping of the 1D
electronic structure because the maxima of P valence bands are
far away from the center of the surface Brillouin zone (SBZ).
As a result, 1D bands of chains with different orientations only partially overlap in {\bf k}-space.

Figs. 2(a--d) report a series of constant energy surfaces (CES) of P chains on Ag(111)
extracted from full photoemission mapping I(E,{\it k}$_{x}$,{\it k}$_{y}$)
for different binding energies.
Circular-like contours labeled as {\it Ag-sp} correspond to 
nearly spherical 3D bulk \textit{sp}-bands of Ag and surface resonances related to them  \cite{Krivenkov-C60-Au111}. 
Electronic states of P chains are seen between {\it Ag-sp} contours
around the $\overline{\rm K}$ point of the Ag(111) SBZ (\pntK) and are labeled as {\it P}. They evolve from a cloverleaf-like pattern at $E_B$=2.26~eV,
through crossing straight double lines at $E_B$=1.76~eV, and to
the triangle of single straight lines at $E_B$=1.46~eV. These 
lines belong to the 1D band structure of P chains. Every side of the 
triangle is emitted from one of the chain orientations, as
sketched in Fig. 3(a).
The overall ARPES dispersion of the band {\it P} along the direction \dirGK\
of Ag(111) SBZ [trace T$_1$ in Fig. 2(d)], which corresponds to the direction along the chains 
of one of the domains, is depicted in Fig. 2(e) and emphasized by a dashed red line.
It has several valleys and shows a pronounced dispersion in a wide energy 
range between $E_B$=3~eV and 1.5~eV. A second band {\it P'} (marked by a dashed blue line) 
is a contribution from two other rotated domains sliced at an angle to a high symmetry direction (therefore, it appears stretched along the k$_\parallel$ axis, as compared to P).

The dispersion in Fig. 2(e) reveals the top of chain band {\it P} at $E_B$$\sim$1.5~eV.
This shows that the intrinsic band gap of phosphorene chains is not less than $E_g$$\sim$1.5~eV, in agreement with our DFT calculations in Fig. 4.

The 1D character of the P band becomes evident when one measures its dispersion 
in the direction perpendicular to the P chains. The ARPES slice along the straight 
line seen in the CES [trace T$_2$ in Fig. 2(d)] is shown in Fig. 2(f). It reveals
perfectly flat slices of the {\it P} band, dispersionless within 20 meV. 
Signals from other rotational domains ({\it P'})
are also seen there.
The band seen at $E_B$$\sim$3.5 eV (white arrows) can be attributed to lower lying P band, which is 
also predicted by our DFT calculations [see Fig. 4(d)].

A sketch of the 1D band of P chains is shown in Fig.~3(b). The model of band
superposition for three orientational domains is displayed in 
Fig. 3(c). This model explains ARPES observations near the \pntK\
point of the Ag(111) SBZ, including the cloverleaf-like pattern at higher {\it E$_{B}$} and triangular 
crossings of double and single straight lines at lower $E_B$. The cross-section on the left side of Fig.~3(c) closely resembles Fig. 2(f). Additional APRES characterization of this superpositional band structure is reported 
in the supporting information \cite{Supplement}.

We also performed a density functional theory (DFT) study of the P chains on Ag(111).
It was assumed that chains are parallel to each other and form a periodic array.
This phase corresponds to the high-phosphorus-coverage sample
used for ARPES measurements and reflects the structure of a single 
rotational domain \cite{BlueP-Zhang-NatComm-2021}.
For the model which reproduces the experimentally synthesized chains resolved by our LEED and STM
[(2$\times$3) relative to Ag(111), Fig. 4(a)], the electronic structure of P chains was 
confirmed as purely 1D. The overall band structure
comprising P and Ag states along the direction \dirGK\ [$\parallel$ to chains] 
and \dirGM\ [$\perp$ to chains] is shown in Figs.~4(b,c). A projection on P 
atoms is presented in Figs.~4(d,e). Good qualitative agreement with the ARPES experiment is seen, including the
dispersionless character of the P chain band perpendicular to the chains direction
due to confinement.

Furthermore, by DFT we studied how the band structure of the P chains
within the array depends on the lateral separation {\it d} between them. 
While the experimentally realized $d=7.6$~\AA\
[Ag(111)/P-(2$\times$3)] hosts a 1D P band, a 
denser simulated array with $d=5.1$~\AA\ [Ag(111)/P-$\smqty(2 & 0 \\ -1 & 2)$] 
triggers inter-chain interaction and causes a 1D$\rightarrow$2D transition in
the electronic structure of the array. The dispersions of the P state calculated for the experimentally synthesized ($d=7.6$~\AA) 
and for artificial ($d=5.1$~\AA) arrays are compared {\it vis-à-vis} 
in Figs. 5(c) and 5(d), respectively. One sees that in the denser array
the P band is not flat but disperses strongly and even crosses the Fermi level ($E_F$), hence rendering the chains metallic.

Finally, we tested the presence of topological edge states, predicted 
for the armchair configuration \cite{BlueP-Tani-PRB-2022}.
These states should appear as nearly flat bands in the vicinity of $E_F$.
We found no such bands below $E_F$ in ARPES and no visible signs of them in STS. 
The absence of topological states can be attributed to their suppression by confinement.
According to Ref. [\onlinecite{BlueP-Tani-PRB-2022}], the localization of edge
states requires a phosphorene nanoribbon with minimal width of 5$-$6
lattice constants, which is not realized in P chains.

\section{Conclusion}

In summary, we studied phosphorene atomic chains self-assembled on a Ag(111) 
substrate in the armchair configuration. 
Although the chains grow in three equiprobable orientations,
aligned to the \hkl<1 -1 0> family of directions of Ag(111) rotated by 120$^\circ$ to each other,
we show that conventional ARPES can be used for the mapping of their band structure  
because their bands are separated in the angular domain. We have measured the 
dispersion of the chain band and show that it is 1D: 
The band disperses within 1.5 eV along the chain
but remains perfectly flat and dispersionless in the direction 
perpendicular to the chain due to lateral confinement. 
Our DFT calculations confirm this result and 
predict a 1D$\rightarrow$2D transition in the electronic 
structure of the chain array when the lateral separation between
the chains is reduced. This transition also changes the band structure
of the chains from semiconducting (with intrinsic band gap $E_g\geq1.5$ eV) to metallic.

\section{Acknowledgments}
This work was supported by Impuls- und Vernetzungsfonds der
Helmholtz-Gemeinschaft under grant no. HRSF-0067.

\section{Conflict of interest}

The authors declare no conflict of interest.

\bibliography{P1DAg111}

\begin{thebibliography}{36}%
\makeatletter
\providecommand \@ifxundefined [1]{%
 \@ifx{#1\undefined}
}%
\providecommand \@ifnum [1]{%
 \ifnum #1\expandafter \@firstoftwo
 \else \expandafter \@secondoftwo
 \fi
}%
\providecommand \@ifx [1]{%
 \ifx #1\expandafter \@firstoftwo
 \else \expandafter \@secondoftwo
 \fi
}%
\providecommand \natexlab [1]{#1}%
\providecommand \enquote  [1]{``#1''}%
\providecommand \bibnamefont  [1]{#1}%
\providecommand \bibfnamefont [1]{#1}%
\providecommand \citenamefont [1]{#1}%
\providecommand \href@noop [0]{\@secondoftwo}%
\providecommand \href [0]{\begingroup \@sanitize@url \@href}%
\providecommand \@href[1]{\@@startlink{#1}\@@href}%
\providecommand \@@href[1]{\endgroup#1\@@endlink}%
\providecommand \@sanitize@url [0]{\catcode `\\12\catcode `\$12\catcode
  `\&12\catcode `\#12\catcode `\^12\catcode `\_12\catcode `\%12\relax}%
\providecommand \@@startlink[1]{}%
\providecommand \@@endlink[0]{}%
\providecommand \url  [0]{\begingroup\@sanitize@url \@url }%
\providecommand \@url [1]{\endgroup\@href {#1}{\urlprefix }}%
\providecommand \urlprefix  [0]{URL }%
\providecommand \Eprint [0]{\href }%
\providecommand \doibase [0]{https://doi.org/}%
\providecommand \selectlanguage [0]{\@gobble}%
\providecommand \bibinfo  [0]{\@secondoftwo}%
\providecommand \bibfield  [0]{\@secondoftwo}%
\providecommand \translation [1]{[#1]}%
\providecommand \BibitemOpen [0]{}%
\providecommand \bibitemStop [0]{}%
\providecommand \bibitemNoStop [0]{.\EOS\space}%
\providecommand \EOS [0]{\spacefactor3000\relax}%
\providecommand \BibitemShut  [1]{\csname bibitem#1\endcsname}%
\let\auto@bib@innerbib\@empty
\bibitem [{\citenamefont {Reis}\ \emph {et~al.}(2017)\citenamefont {Reis},
  \citenamefont {Li}, \citenamefont {Dudy}, \citenamefont {Bauernfeind},
  \citenamefont {Glass}, \citenamefont {Hanke}, \citenamefont {Thomale},
  \citenamefont {Schäfer},\ and\ \citenamefont
  {Claessen}}]{Bismuthene_Reis2017}%
  \BibitemOpen
  \bibfield  {author} {\bibinfo {author} {\bibfnamefont {F.}~\bibnamefont
  {Reis}}, \bibinfo {author} {\bibfnamefont {G.}~\bibnamefont {Li}}, \bibinfo
  {author} {\bibfnamefont {L.}~\bibnamefont {Dudy}}, \bibinfo {author}
  {\bibfnamefont {M.}~\bibnamefont {Bauernfeind}}, \bibinfo {author}
  {\bibfnamefont {S.}~\bibnamefont {Glass}}, \bibinfo {author} {\bibfnamefont
  {W.}~\bibnamefont {Hanke}}, \bibinfo {author} {\bibfnamefont
  {R.}~\bibnamefont {Thomale}}, \bibinfo {author} {\bibfnamefont
  {J.}~\bibnamefont {Schäfer}},\ and\ \bibinfo {author} {\bibfnamefont
  {R.}~\bibnamefont {Claessen}},\ }\bibfield  {title} {\bibinfo {title}
  {Bismuthene on a {SiC} substrate: A candidate for a high-temperature quantum
  spin {Hall} material},\ }\href {https://doi.org/10.1126/science.aai8142}
  {\bibfield  {journal} {\bibinfo  {journal} {Science}\ }\textbf {\bibinfo
  {volume} {357}},\ \bibinfo {pages} {287} (\bibinfo {year}
  {2017})}\BibitemShut {NoStop}%
\bibitem [{\citenamefont {Shah}\ \emph {et~al.}(2020)\citenamefont {Shah},
  \citenamefont {Wang}, \citenamefont {Sohail},\ and\ \citenamefont
  {Uhrberg}}]{Shah2020}%
  \BibitemOpen
  \bibfield  {author} {\bibinfo {author} {\bibfnamefont {J.}~\bibnamefont
  {Shah}}, \bibinfo {author} {\bibfnamefont {W.}~\bibnamefont {Wang}}, \bibinfo
  {author} {\bibfnamefont {H.~M.}\ \bibnamefont {Sohail}},\ and\ \bibinfo
  {author} {\bibfnamefont {R.~I.~G.}\ \bibnamefont {Uhrberg}},\ }\bibfield
  {title} {\bibinfo {title} {Experimental evidence of monolayer arsenene: an
  exotic {2D} semiconducting material},\ }\href
  {https://doi.org/10.1088/2053-1583/ab64fb} {\bibfield  {journal} {\bibinfo
  {journal} {2D Mater.}\ }\textbf {\bibinfo {volume} {7}},\ \bibinfo {pages}
  {025013} (\bibinfo {year} {2020})}\BibitemShut {NoStop}%
\bibitem [{\citenamefont {Ling}\ \emph {et~al.}(2015)\citenamefont {Ling},
  \citenamefont {Wang}, \citenamefont {Huang}, \citenamefont {Xia},\ and\
  \citenamefont {Dresselhaus}}]{BlackP-Dresselhaus-2015}%
  \BibitemOpen
  \bibfield  {author} {\bibinfo {author} {\bibfnamefont {X.}~\bibnamefont
  {Ling}}, \bibinfo {author} {\bibfnamefont {H.}~\bibnamefont {Wang}}, \bibinfo
  {author} {\bibfnamefont {S.}~\bibnamefont {Huang}}, \bibinfo {author}
  {\bibfnamefont {F.}~\bibnamefont {Xia}},\ and\ \bibinfo {author}
  {\bibfnamefont {M.~S.}\ \bibnamefont {Dresselhaus}},\ }\bibfield  {title}
  {\bibinfo {title} {The renaissance of black phosphorus},\ }\href
  {https://doi.org/10.1073/pnas.1416581112} {\bibfield  {journal} {\bibinfo
  {journal} {PNAS}\ }\textbf {\bibinfo {volume} {112}},\ \bibinfo {pages}
  {4523} (\bibinfo {year} {2015})}\BibitemShut {NoStop}%
\bibitem [{\citenamefont {Han}\ \emph {et~al.}(2014)\citenamefont {Han},
  \citenamefont {Yao}, \citenamefont {Bai}, \citenamefont {Miao}, \citenamefont
  {Zhu}, \citenamefont {Guan}, \citenamefont {Wang}, \citenamefont {Gao},
  \citenamefont {Liu}, \citenamefont {Qian}, \citenamefont {Liu},\ and\
  \citenamefont {Jia}}]{BlackP-Han-2014}%
  \BibitemOpen
  \bibfield  {author} {\bibinfo {author} {\bibfnamefont {C.~Q.}\ \bibnamefont
  {Han}}, \bibinfo {author} {\bibfnamefont {M.~Y.}\ \bibnamefont {Yao}},
  \bibinfo {author} {\bibfnamefont {X.~X.}\ \bibnamefont {Bai}}, \bibinfo
  {author} {\bibfnamefont {L.}~\bibnamefont {Miao}}, \bibinfo {author}
  {\bibfnamefont {F.}~\bibnamefont {Zhu}}, \bibinfo {author} {\bibfnamefont
  {D.~D.}\ \bibnamefont {Guan}}, \bibinfo {author} {\bibfnamefont
  {S.}~\bibnamefont {Wang}}, \bibinfo {author} {\bibfnamefont {C.~L.}\
  \bibnamefont {Gao}}, \bibinfo {author} {\bibfnamefont {C.}~\bibnamefont
  {Liu}}, \bibinfo {author} {\bibfnamefont {D.}~\bibnamefont {Qian}}, \bibinfo
  {author} {\bibfnamefont {Y.}~\bibnamefont {Liu}},\ and\ \bibinfo {author}
  {\bibfnamefont {J.-f.}\ \bibnamefont {Jia}},\ }\bibfield  {title} {\bibinfo
  {title} {Electronic structure of black phosphorus studied by angle-resolved
  photoemission spectroscopy},\ }\href
  {https://doi.org/10.1103/physrevb.90.085101} {\bibfield  {journal} {\bibinfo
  {journal} {Phys. Rev. B}\ }\textbf {\bibinfo {volume} {90}},\ \bibinfo
  {pages} {085101} (\bibinfo {year} {2014})}\BibitemShut {NoStop}%
\bibitem [{\citenamefont {Xia}\ \emph {et~al.}(2019)\citenamefont {Xia},
  \citenamefont {Wang}, \citenamefont {Hwang}, \citenamefont {Neto},\ and\
  \citenamefont {Yang}}]{BlackP-Xia-2019}%
  \BibitemOpen
  \bibfield  {author} {\bibinfo {author} {\bibfnamefont {F.}~\bibnamefont
  {Xia}}, \bibinfo {author} {\bibfnamefont {H.}~\bibnamefont {Wang}}, \bibinfo
  {author} {\bibfnamefont {J.~C.~M.}\ \bibnamefont {Hwang}}, \bibinfo {author}
  {\bibfnamefont {A.~H.~C.}\ \bibnamefont {Neto}},\ and\ \bibinfo {author}
  {\bibfnamefont {L.}~\bibnamefont {Yang}},\ }\bibfield  {title} {\bibinfo
  {title} {Black phosphorus and its isoelectronic materials},\ }\href
  {https://doi.org/10.1038/s42254-019-0043-5} {\bibfield  {journal} {\bibinfo
  {journal} {Nat. Rev. Physics}\ }\textbf {\bibinfo {volume} {1}},\ \bibinfo
  {pages} {306} (\bibinfo {year} {2019})}\BibitemShut {NoStop}%
\bibitem [{\citenamefont {Golias}\ \emph {et~al.}(2016)\citenamefont {Golias},
  \citenamefont {Krivenkov},\ and\ \citenamefont
  {S{\'{a}}nchez-Barriga}}]{BlackP-Golias-2016}%
  \BibitemOpen
  \bibfield  {author} {\bibinfo {author} {\bibfnamefont {E.}~\bibnamefont
  {Golias}}, \bibinfo {author} {\bibfnamefont {M.}~\bibnamefont {Krivenkov}},\
  and\ \bibinfo {author} {\bibfnamefont {J.}~\bibnamefont
  {S{\'{a}}nchez-Barriga}},\ }\bibfield  {title} {\bibinfo {title}
  {Disentangling bulk from surface contributions in the electronic structure of
  black phosphorus},\ }\href {https://doi.org/10.1103/physrevb.93.075207}
  {\bibfield  {journal} {\bibinfo  {journal} {Phys. Rev. B}\ }\textbf {\bibinfo
  {volume} {93}},\ \bibinfo {pages} {075207} (\bibinfo {year}
  {2016})}\BibitemShut {NoStop}%
\bibitem [{\citenamefont {Kim}\ \emph {et~al.}(2015)\citenamefont {Kim},
  \citenamefont {Baik}, \citenamefont {Ryu}, \citenamefont {Sohn},
  \citenamefont {Park}, \citenamefont {Park}, \citenamefont {Denlinger},
  \citenamefont {Yi}, \citenamefont {Choi},\ and\ \citenamefont
  {Kim}}]{Kim-Science-2015}%
  \BibitemOpen
  \bibfield  {author} {\bibinfo {author} {\bibfnamefont {J.}~\bibnamefont
  {Kim}}, \bibinfo {author} {\bibfnamefont {S.~S.}\ \bibnamefont {Baik}},
  \bibinfo {author} {\bibfnamefont {S.~H.}\ \bibnamefont {Ryu}}, \bibinfo
  {author} {\bibfnamefont {Y.}~\bibnamefont {Sohn}}, \bibinfo {author}
  {\bibfnamefont {S.}~\bibnamefont {Park}}, \bibinfo {author} {\bibfnamefont
  {B.-G.}\ \bibnamefont {Park}}, \bibinfo {author} {\bibfnamefont
  {J.}~\bibnamefont {Denlinger}}, \bibinfo {author} {\bibfnamefont
  {Y.}~\bibnamefont {Yi}}, \bibinfo {author} {\bibfnamefont {H.~J.}\
  \bibnamefont {Choi}},\ and\ \bibinfo {author} {\bibfnamefont {K.~S.}\
  \bibnamefont {Kim}},\ }\bibfield  {title} {\bibinfo {title} {Observation of
  tunable band gap and anisotropic {Dirac} semimetal state in black
  phosphorus},\ }\href {https://doi.org/10.1126/science.aaa6486} {\bibfield
  {journal} {\bibinfo  {journal} {Science}\ }\textbf {\bibinfo {volume}
  {349}},\ \bibinfo {pages} {723} (\bibinfo {year} {2015})}\BibitemShut
  {NoStop}%
\bibitem [{\citenamefont {Li}\ \emph {et~al.}(2014)\citenamefont {Li},
  \citenamefont {Yu}, \citenamefont {Ye}, \citenamefont {Ge}, \citenamefont
  {Ou}, \citenamefont {Wu}, \citenamefont {Feng}, \citenamefont {Chen},\ and\
  \citenamefont {Zhang}}]{Li2014}%
  \BibitemOpen
  \bibfield  {author} {\bibinfo {author} {\bibfnamefont {L.}~\bibnamefont
  {Li}}, \bibinfo {author} {\bibfnamefont {Y.}~\bibnamefont {Yu}}, \bibinfo
  {author} {\bibfnamefont {G.~J.}\ \bibnamefont {Ye}}, \bibinfo {author}
  {\bibfnamefont {Q.}~\bibnamefont {Ge}}, \bibinfo {author} {\bibfnamefont
  {X.}~\bibnamefont {Ou}}, \bibinfo {author} {\bibfnamefont {H.}~\bibnamefont
  {Wu}}, \bibinfo {author} {\bibfnamefont {D.}~\bibnamefont {Feng}}, \bibinfo
  {author} {\bibfnamefont {X.~H.}\ \bibnamefont {Chen}},\ and\ \bibinfo
  {author} {\bibfnamefont {Y.}~\bibnamefont {Zhang}},\ }\bibfield  {title}
  {\bibinfo {title} {Black phosphorus field-effect transistors},\ }\href
  {https://doi.org/10.1038/nnano.2014.35} {\bibfield  {journal} {\bibinfo
  {journal} {Nat. Nanotechnol.}\ }\textbf {\bibinfo {volume} {9}},\ \bibinfo
  {pages} {372} (\bibinfo {year} {2014})}\BibitemShut {NoStop}%
\bibitem [{\citenamefont {Xia}\ \emph {et~al.}(2014)\citenamefont {Xia},
  \citenamefont {Wang},\ and\ \citenamefont {Jia}}]{Xia2014}%
  \BibitemOpen
  \bibfield  {author} {\bibinfo {author} {\bibfnamefont {F.}~\bibnamefont
  {Xia}}, \bibinfo {author} {\bibfnamefont {H.}~\bibnamefont {Wang}},\ and\
  \bibinfo {author} {\bibfnamefont {Y.}~\bibnamefont {Jia}},\ }\bibfield
  {title} {\bibinfo {title} {Rediscovering black phosphorus as an anisotropic
  layered material for optoelectronics and electronics},\ }\href
  {https://doi.org/10.1038/ncomms5458} {\bibfield  {journal} {\bibinfo
  {journal} {Nat. Commun.}\ }\textbf {\bibinfo {volume} {5}},\ \bibinfo {pages}
  {4458} (\bibinfo {year} {2014})}\BibitemShut {NoStop}%
\bibitem [{\citenamefont {Zhang}\ \emph {et~al.}(2016)\citenamefont {Zhang},
  \citenamefont {Zhao}, \citenamefont {Han}, \citenamefont {Wang},
  \citenamefont {Zhong}, \citenamefont {Sun}, \citenamefont {Guo},
  \citenamefont {Zhou}, \citenamefont {Gu}, \citenamefont {Yuan}, \citenamefont
  {Li},\ and\ \citenamefont {Chen}}]{BlueP-Zhang-Nanoletters-2016}%
  \BibitemOpen
  \bibfield  {author} {\bibinfo {author} {\bibfnamefont {J.~L.}\ \bibnamefont
  {Zhang}}, \bibinfo {author} {\bibfnamefont {S.}~\bibnamefont {Zhao}},
  \bibinfo {author} {\bibfnamefont {C.}~\bibnamefont {Han}}, \bibinfo {author}
  {\bibfnamefont {Z.}~\bibnamefont {Wang}}, \bibinfo {author} {\bibfnamefont
  {S.}~\bibnamefont {Zhong}}, \bibinfo {author} {\bibfnamefont
  {S.}~\bibnamefont {Sun}}, \bibinfo {author} {\bibfnamefont {R.}~\bibnamefont
  {Guo}}, \bibinfo {author} {\bibfnamefont {X.}~\bibnamefont {Zhou}}, \bibinfo
  {author} {\bibfnamefont {C.~D.}\ \bibnamefont {Gu}}, \bibinfo {author}
  {\bibfnamefont {K.~D.}\ \bibnamefont {Yuan}}, \bibinfo {author}
  {\bibfnamefont {Z.}~\bibnamefont {Li}},\ and\ \bibinfo {author}
  {\bibfnamefont {W.}~\bibnamefont {Chen}},\ }\bibfield  {title} {\bibinfo
  {title} {Epitaxial growth of single layer blue phosphorus: A new phase of
  two-dimensional phosphorus},\ }\href
  {https://doi.org/10.1021/acs.nanolett.6b01459} {\bibfield  {journal}
  {\bibinfo  {journal} {Nano Lett.}\ }\textbf {\bibinfo {volume} {16}},\
  \bibinfo {pages} {4903} (\bibinfo {year} {2016})}\BibitemShut {NoStop}%
\bibitem [{\citenamefont {Zhang}\ \emph {et~al.}(2018)\citenamefont {Zhang},
  \citenamefont {Enriquez}, \citenamefont {Tong}, \citenamefont {Bendounan},
  \citenamefont {Kara}, \citenamefont {Seitsonen}, \citenamefont {Mayne},
  \citenamefont {Dujardin},\ and\ \citenamefont
  {Oughaddou}}]{BlueP-Zhang-Small-2018}%
  \BibitemOpen
  \bibfield  {author} {\bibinfo {author} {\bibfnamefont {W.}~\bibnamefont
  {Zhang}}, \bibinfo {author} {\bibfnamefont {H.}~\bibnamefont {Enriquez}},
  \bibinfo {author} {\bibfnamefont {Y.}~\bibnamefont {Tong}}, \bibinfo {author}
  {\bibfnamefont {A.}~\bibnamefont {Bendounan}}, \bibinfo {author}
  {\bibfnamefont {A.}~\bibnamefont {Kara}}, \bibinfo {author} {\bibfnamefont
  {A.~P.}\ \bibnamefont {Seitsonen}}, \bibinfo {author} {\bibfnamefont {A.~J.}\
  \bibnamefont {Mayne}}, \bibinfo {author} {\bibfnamefont {G.}~\bibnamefont
  {Dujardin}},\ and\ \bibinfo {author} {\bibfnamefont {H.}~\bibnamefont
  {Oughaddou}},\ }\bibfield  {title} {\bibinfo {title} {Epitaxial synthesis of
  blue phosphorene},\ }\href {https://doi.org/10.1002/smll.201804066}
  {\bibfield  {journal} {\bibinfo  {journal} {Small}\ }\textbf {\bibinfo
  {volume} {14}},\ \bibinfo {pages} {1804066} (\bibinfo {year}
  {2018})}\BibitemShut {NoStop}%
\bibitem [{\citenamefont {Schaal}\ \emph {et~al.}(2021)\citenamefont {Schaal},
  \citenamefont {Picker}, \citenamefont {Otto}, \citenamefont {Gruenewald},
  \citenamefont {Forker},\ and\ \citenamefont {Fritz}}]{PAu100_Schaal2021}%
  \BibitemOpen
  \bibfield  {author} {\bibinfo {author} {\bibfnamefont {M.}~\bibnamefont
  {Schaal}}, \bibinfo {author} {\bibfnamefont {J.}~\bibnamefont {Picker}},
  \bibinfo {author} {\bibfnamefont {F.}~\bibnamefont {Otto}}, \bibinfo {author}
  {\bibfnamefont {M.}~\bibnamefont {Gruenewald}}, \bibinfo {author}
  {\bibfnamefont {R.}~\bibnamefont {Forker}},\ and\ \bibinfo {author}
  {\bibfnamefont {T.}~\bibnamefont {Fritz}},\ }\bibfield  {title} {\bibinfo
  {title} {An alternative route towards the fabrication of {2D} blue
  phosphorene},\ }\href {https://doi.org/10.1088/1361-648x/ac1dde} {\bibfield
  {journal} {\bibinfo  {journal} {J. Phys.: Condens. Matter}\ }\textbf
  {\bibinfo {volume} {33}},\ \bibinfo {pages} {485002} (\bibinfo {year}
  {2021})}\BibitemShut {NoStop}%
\bibitem [{\citenamefont {Yang}\ \emph {et~al.}(2020)\citenamefont {Yang},
  \citenamefont {Hu}, \citenamefont {Wang}, \citenamefont {Cheng},
  \citenamefont {Chen},\ and\ \citenamefont {Wu}}]{BlueP_Ag111_Yang2020}%
  \BibitemOpen
  \bibfield  {author} {\bibinfo {author} {\bibfnamefont {S.}~\bibnamefont
  {Yang}}, \bibinfo {author} {\bibfnamefont {Z.}~\bibnamefont {Hu}}, \bibinfo
  {author} {\bibfnamefont {W.}~\bibnamefont {Wang}}, \bibinfo {author}
  {\bibfnamefont {P.}~\bibnamefont {Cheng}}, \bibinfo {author} {\bibfnamefont
  {L.}~\bibnamefont {Chen}},\ and\ \bibinfo {author} {\bibfnamefont
  {K.}~\bibnamefont {Wu}},\ }\bibfield  {title} {\bibinfo {title} {Regular
  arrangement of two-dimensional clusters of {Blue} {Phosphorene} on
  {Ag(111)}},\ }\href {https://doi.org/10.1088/0256-307x/37/9/096803}
  {\bibfield  {journal} {\bibinfo  {journal} {Chin. Phys. Lett.}\ }\textbf
  {\bibinfo {volume} {37}},\ \bibinfo {pages} {096803} (\bibinfo {year}
  {2020})}\BibitemShut {NoStop}%
\bibitem [{\citenamefont {Zhang}\ \emph {et~al.}(2022)\citenamefont {Zhang},
  \citenamefont {Dong}, \citenamefont {Xu}, \citenamefont {Xia}, \citenamefont
  {Ho}, \citenamefont {Xu},\ and\ \citenamefont {Xie}}]{Zhang2022}%
  \BibitemOpen
  \bibfield  {author} {\bibinfo {author} {\bibfnamefont {J.}~\bibnamefont
  {Zhang}}, \bibinfo {author} {\bibfnamefont {X.}~\bibnamefont {Dong}},
  \bibinfo {author} {\bibfnamefont {S.}~\bibnamefont {Xu}}, \bibinfo {author}
  {\bibfnamefont {Y.}~\bibnamefont {Xia}}, \bibinfo {author} {\bibfnamefont
  {W.}~\bibnamefont {Ho}}, \bibinfo {author} {\bibfnamefont {H.}~\bibnamefont
  {Xu}},\ and\ \bibinfo {author} {\bibfnamefont {M.}~\bibnamefont {Xie}},\
  }\bibfield  {title} {\bibinfo {title} {Metal-phosphorus network on
  {Pt(111)}},\ }\href {https://doi.org/10.1088/2053-1583/ac780c} {\bibfield
  {journal} {\bibinfo  {journal} {2D Mater.}\ }\textbf {\bibinfo {volume}
  {9}},\ \bibinfo {pages} {045002} (\bibinfo {year} {2022})}\BibitemShut
  {NoStop}%
\bibitem [{\citenamefont {Zhuang}\ \emph {et~al.}(2018)\citenamefont {Zhuang},
  \citenamefont {Liu}, \citenamefont {Gao}, \citenamefont {Liu}, \citenamefont
  {Feng}, \citenamefont {Xu}, \citenamefont {Wang}, \citenamefont {Zhao},
  \citenamefont {Dou}, \citenamefont {Hu},\ and\ \citenamefont
  {Du}}]{BlueP-Zhuang-ACSNano-2018}%
  \BibitemOpen
  \bibfield  {author} {\bibinfo {author} {\bibfnamefont {J.}~\bibnamefont
  {Zhuang}}, \bibinfo {author} {\bibfnamefont {C.}~\bibnamefont {Liu}},
  \bibinfo {author} {\bibfnamefont {Q.}~\bibnamefont {Gao}}, \bibinfo {author}
  {\bibfnamefont {Y.}~\bibnamefont {Liu}}, \bibinfo {author} {\bibfnamefont
  {H.}~\bibnamefont {Feng}}, \bibinfo {author} {\bibfnamefont {X.}~\bibnamefont
  {Xu}}, \bibinfo {author} {\bibfnamefont {J.}~\bibnamefont {Wang}}, \bibinfo
  {author} {\bibfnamefont {J.}~\bibnamefont {Zhao}}, \bibinfo {author}
  {\bibfnamefont {S.~X.}\ \bibnamefont {Dou}}, \bibinfo {author} {\bibfnamefont
  {Z.}~\bibnamefont {Hu}},\ and\ \bibinfo {author} {\bibfnamefont
  {Y.}~\bibnamefont {Du}},\ }\bibfield  {title} {\bibinfo {title} {Band gap
  modulated by electronic superlattice in blue phosphorene},\ }\href
  {https://doi.org/10.1021/acsnano.8b02953} {\bibfield  {journal} {\bibinfo
  {journal} {{ACS} Nano}\ }\textbf {\bibinfo {volume} {12}},\ \bibinfo {pages}
  {5059} (\bibinfo {year} {2018})}\BibitemShut {NoStop}%
\bibitem [{\citenamefont {Golias}\ \emph {et~al.}(2018)\citenamefont {Golias},
  \citenamefont {Krivenkov}, \citenamefont {Varykhalov}, \citenamefont
  {S{\'{a}}nchez-Barriga},\ and\ \citenamefont
  {Rader}}]{BlueP-Golias-Nanoletters-2018}%
  \BibitemOpen
  \bibfield  {author} {\bibinfo {author} {\bibfnamefont {E.}~\bibnamefont
  {Golias}}, \bibinfo {author} {\bibfnamefont {M.}~\bibnamefont {Krivenkov}},
  \bibinfo {author} {\bibfnamefont {A.}~\bibnamefont {Varykhalov}}, \bibinfo
  {author} {\bibfnamefont {J.}~\bibnamefont {S{\'{a}}nchez-Barriga}},\ and\
  \bibinfo {author} {\bibfnamefont {O.}~\bibnamefont {Rader}},\ }\bibfield
  {title} {\bibinfo {title} {Band renormalization of blue phosphorus on
  {Au(111)}},\ }\href {https://doi.org/10.1021/acs.nanolett.8b01305} {\bibfield
   {journal} {\bibinfo  {journal} {Nano Lett.}\ }\textbf {\bibinfo {volume}
  {18}},\ \bibinfo {pages} {6672} (\bibinfo {year} {2018})}\BibitemShut
  {NoStop}%
\bibitem [{\citenamefont {Wu}\ \emph {et~al.}(2015)\citenamefont {Wu},
  \citenamefont {Shen}, \citenamefont {Yang}, \citenamefont {Cai},
  \citenamefont {Huang},\ and\ \citenamefont {Feng}}]{Wu2015}%
  \BibitemOpen
  \bibfield  {author} {\bibinfo {author} {\bibfnamefont {Q.}~\bibnamefont
  {Wu}}, \bibinfo {author} {\bibfnamefont {L.}~\bibnamefont {Shen}}, \bibinfo
  {author} {\bibfnamefont {M.}~\bibnamefont {Yang}}, \bibinfo {author}
  {\bibfnamefont {Y.}~\bibnamefont {Cai}}, \bibinfo {author} {\bibfnamefont
  {Z.}~\bibnamefont {Huang}},\ and\ \bibinfo {author} {\bibfnamefont {Y.~P.}\
  \bibnamefont {Feng}},\ }\bibfield  {title} {\bibinfo {title} {Electronic and
  transport properties of phosphorene nanoribbons},\ }\href
  {https://doi.org/10.1103/PhysRevB.92.035436} {\bibfield  {journal} {\bibinfo
  {journal} {Phys. Rev. B}\ }\textbf {\bibinfo {volume} {92}},\ \bibinfo
  {pages} {035436} (\bibinfo {year} {2015})}\BibitemShut {NoStop}%
\bibitem [{\citenamefont {Taghizadeh~Sisakht}\ \emph
  {et~al.}(2016)\citenamefont {Taghizadeh~Sisakht}, \citenamefont {Fazileh},
  \citenamefont {Zare}, \citenamefont {Zarenia},\ and\ \citenamefont
  {Peeters}}]{Sisakht2016}%
  \BibitemOpen
  \bibfield  {author} {\bibinfo {author} {\bibfnamefont {E.}~\bibnamefont
  {Taghizadeh~Sisakht}}, \bibinfo {author} {\bibfnamefont {F.}~\bibnamefont
  {Fazileh}}, \bibinfo {author} {\bibfnamefont {M.~H.}\ \bibnamefont {Zare}},
  \bibinfo {author} {\bibfnamefont {M.}~\bibnamefont {Zarenia}},\ and\ \bibinfo
  {author} {\bibfnamefont {F.~M.}\ \bibnamefont {Peeters}},\ }\bibfield
  {title} {\bibinfo {title} {Strain-induced topological phase transition in
  phosphorene and in phosphorene nanoribbons},\ }\href
  {https://doi.org/10.1103/PhysRevB.94.085417} {\bibfield  {journal} {\bibinfo
  {journal} {Phys. Rev. B}\ }\textbf {\bibinfo {volume} {94}},\ \bibinfo
  {pages} {085417} (\bibinfo {year} {2016})}\BibitemShut {NoStop}%
\bibitem [{\citenamefont {Cai}\ \emph {et~al.}(2010)\citenamefont {Cai},
  \citenamefont {Ruffieux}, \citenamefont {Jaafar}, \citenamefont {Bieri},
  \citenamefont {Braun}, \citenamefont {Blankenburg}, \citenamefont {Muoth},
  \citenamefont {Seitsonen}, \citenamefont {Saleh}, \citenamefont {Feng},
  \citenamefont {Müllen},\ and\ \citenamefont {Fasel}}]{Cai2010}%
  \BibitemOpen
  \bibfield  {author} {\bibinfo {author} {\bibfnamefont {J.}~\bibnamefont
  {Cai}}, \bibinfo {author} {\bibfnamefont {P.}~\bibnamefont {Ruffieux}},
  \bibinfo {author} {\bibfnamefont {R.}~\bibnamefont {Jaafar}}, \bibinfo
  {author} {\bibfnamefont {M.}~\bibnamefont {Bieri}}, \bibinfo {author}
  {\bibfnamefont {T.}~\bibnamefont {Braun}}, \bibinfo {author} {\bibfnamefont
  {S.}~\bibnamefont {Blankenburg}}, \bibinfo {author} {\bibfnamefont
  {M.}~\bibnamefont {Muoth}}, \bibinfo {author} {\bibfnamefont {A.~P.}\
  \bibnamefont {Seitsonen}}, \bibinfo {author} {\bibfnamefont {M.}~\bibnamefont
  {Saleh}}, \bibinfo {author} {\bibfnamefont {X.}~\bibnamefont {Feng}},
  \bibinfo {author} {\bibfnamefont {K.}~\bibnamefont {Müllen}},\ and\ \bibinfo
  {author} {\bibfnamefont {R.}~\bibnamefont {Fasel}},\ }\bibfield  {title}
  {\bibinfo {title} {Atomically precise bottom-up fabrication of graphene
  nanoribbons},\ }\href {https://doi.org/10.1038/nature09211} {\bibfield
  {journal} {\bibinfo  {journal} {Nature}\ }\textbf {\bibinfo {volume} {466}},\
  \bibinfo {pages} {470} (\bibinfo {year} {2010})}\BibitemShut {NoStop}%
\bibitem [{\citenamefont {Watts}\ \emph {et~al.}(2019)\citenamefont {Watts},
  \citenamefont {Picco}, \citenamefont {Russell-Pavier}, \citenamefont
  {Cullen}, \citenamefont {Miller}, \citenamefont {Bartuś}, \citenamefont
  {Payton}, \citenamefont {Skipper}, \citenamefont {Tileli},\ and\
  \citenamefont {Howard}}]{Watts2019}%
  \BibitemOpen
  \bibfield  {author} {\bibinfo {author} {\bibfnamefont {M.~C.}\ \bibnamefont
  {Watts}}, \bibinfo {author} {\bibfnamefont {L.}~\bibnamefont {Picco}},
  \bibinfo {author} {\bibfnamefont {F.~S.}\ \bibnamefont {Russell-Pavier}},
  \bibinfo {author} {\bibfnamefont {P.~L.}\ \bibnamefont {Cullen}}, \bibinfo
  {author} {\bibfnamefont {T.~S.}\ \bibnamefont {Miller}}, \bibinfo {author}
  {\bibfnamefont {S.~P.}\ \bibnamefont {Bartuś}}, \bibinfo {author}
  {\bibfnamefont {O.~D.}\ \bibnamefont {Payton}}, \bibinfo {author}
  {\bibfnamefont {N.~T.}\ \bibnamefont {Skipper}}, \bibinfo {author}
  {\bibfnamefont {V.}~\bibnamefont {Tileli}},\ and\ \bibinfo {author}
  {\bibfnamefont {C.~A.}\ \bibnamefont {Howard}},\ }\bibfield  {title}
  {\bibinfo {title} {Production of phosphorene nanoribbons},\ }\href
  {https://doi.org/10.1038/s41586-019-1074-x} {\bibfield  {journal} {\bibinfo
  {journal} {Nature}\ }\textbf {\bibinfo {volume} {568}},\ \bibinfo {pages}
  {216} (\bibinfo {year} {2019})}\BibitemShut {NoStop}%
\bibitem [{\citenamefont {Lim}\ \emph {et~al.}(2022)\citenamefont {Lim},
  \citenamefont {Liu}, \citenamefont {Kim}, \citenamefont {Pu}, \citenamefont
  {Shimizu}, \citenamefont {Endo}, \citenamefont {Nakanishi}, \citenamefont
  {Takenobu},\ and\ \citenamefont {Miyata}}]{Lim2022}%
  \BibitemOpen
  \bibfield  {author} {\bibinfo {author} {\bibfnamefont {H.~E.}\ \bibnamefont
  {Lim}}, \bibinfo {author} {\bibfnamefont {Z.}~\bibnamefont {Liu}}, \bibinfo
  {author} {\bibfnamefont {J.}~\bibnamefont {Kim}}, \bibinfo {author}
  {\bibfnamefont {J.}~\bibnamefont {Pu}}, \bibinfo {author} {\bibfnamefont
  {H.}~\bibnamefont {Shimizu}}, \bibinfo {author} {\bibfnamefont
  {T.}~\bibnamefont {Endo}}, \bibinfo {author} {\bibfnamefont {Y.}~\bibnamefont
  {Nakanishi}}, \bibinfo {author} {\bibfnamefont {T.}~\bibnamefont
  {Takenobu}},\ and\ \bibinfo {author} {\bibfnamefont {Y.}~\bibnamefont
  {Miyata}},\ }\bibfield  {title} {\bibinfo {title} {Nanowire-to-nanoribbon
  conversion in transition-metal chalcogenides: Implications for
  one-dimensional electronics and optoelectronics},\ }\href
  {https://doi.org/10.1021/acsanm.1c03160} {\bibfield  {journal} {\bibinfo
  {journal} {ACS Appl. Nano Mater.}\ }\textbf {\bibinfo {volume} {5}},\
  \bibinfo {pages} {1775} (\bibinfo {year} {2022})}\BibitemShut {NoStop}%
\bibitem [{\citenamefont {Zhang}\ \emph {et~al.}(2021)\citenamefont {Zhang},
  \citenamefont {Enriquez}, \citenamefont {Tong}, \citenamefont {Mayne},
  \citenamefont {Bendounan}, \citenamefont {Smogunov}, \citenamefont {Dappe},
  \citenamefont {Kara}, \citenamefont {Dujardin},\ and\ \citenamefont
  {Oughaddou}}]{BlueP-Zhang-NatComm-2021}%
  \BibitemOpen
  \bibfield  {author} {\bibinfo {author} {\bibfnamefont {W.}~\bibnamefont
  {Zhang}}, \bibinfo {author} {\bibfnamefont {H.}~\bibnamefont {Enriquez}},
  \bibinfo {author} {\bibfnamefont {Y.}~\bibnamefont {Tong}}, \bibinfo {author}
  {\bibfnamefont {A.~J.}\ \bibnamefont {Mayne}}, \bibinfo {author}
  {\bibfnamefont {A.}~\bibnamefont {Bendounan}}, \bibinfo {author}
  {\bibfnamefont {A.}~\bibnamefont {Smogunov}}, \bibinfo {author}
  {\bibfnamefont {Y.~J.}\ \bibnamefont {Dappe}}, \bibinfo {author}
  {\bibfnamefont {A.}~\bibnamefont {Kara}}, \bibinfo {author} {\bibfnamefont
  {G.}~\bibnamefont {Dujardin}},\ and\ \bibinfo {author} {\bibfnamefont
  {H.}~\bibnamefont {Oughaddou}},\ }\bibfield  {title} {\bibinfo {title} {Flat
  epitaxial quasi-{1D} phosphorene chains},\ }\href
  {https://doi.org/10.1038/s41467-021-25262-7} {\bibfield  {journal} {\bibinfo
  {journal} {Nat. Commun.}\ }\textbf {\bibinfo {volume} {12}},\ \bibinfo
  {pages} {5160} (\bibinfo {year} {2021})}\BibitemShut {NoStop}%
\bibitem [{\citenamefont {Segovia}\ \emph {et~al.}(1999)\citenamefont
  {Segovia}, \citenamefont {Purdie}, \citenamefont {Hengsberger},\ and\
  \citenamefont {Baer}}]{Segovia1999}%
  \BibitemOpen
  \bibfield  {author} {\bibinfo {author} {\bibfnamefont {P.}~\bibnamefont
  {Segovia}}, \bibinfo {author} {\bibfnamefont {D.}~\bibnamefont {Purdie}},
  \bibinfo {author} {\bibfnamefont {M.}~\bibnamefont {Hengsberger}},\ and\
  \bibinfo {author} {\bibfnamefont {Y.}~\bibnamefont {Baer}},\ }\bibfield
  {title} {\bibinfo {title} {Observation of spin and charge collective modes in
  one-dimensional metallic chains},\ }\href {https://doi.org/10.1038/990052}
  {\bibfield  {journal} {\bibinfo  {journal} {Nature}\ }\textbf {\bibinfo
  {volume} {402}},\ \bibinfo {pages} {504} (\bibinfo {year}
  {1999})}\BibitemShut {NoStop}%
\bibitem [{\citenamefont {Crain}\ \emph {et~al.}(2003)\citenamefont {Crain},
  \citenamefont {Kirakosian}, \citenamefont {Altmann}, \citenamefont
  {Bromberger}, \citenamefont {Erwin}, \citenamefont {McChesney}, \citenamefont
  {Lin},\ and\ \citenamefont {Himpsel}}]{Crain2003}%
  \BibitemOpen
  \bibfield  {author} {\bibinfo {author} {\bibfnamefont {J.~N.}\ \bibnamefont
  {Crain}}, \bibinfo {author} {\bibfnamefont {A.}~\bibnamefont {Kirakosian}},
  \bibinfo {author} {\bibfnamefont {K.~N.}\ \bibnamefont {Altmann}}, \bibinfo
  {author} {\bibfnamefont {C.}~\bibnamefont {Bromberger}}, \bibinfo {author}
  {\bibfnamefont {S.~C.}\ \bibnamefont {Erwin}}, \bibinfo {author}
  {\bibfnamefont {J.~L.}\ \bibnamefont {McChesney}}, \bibinfo {author}
  {\bibfnamefont {J.-L.}\ \bibnamefont {Lin}},\ and\ \bibinfo {author}
  {\bibfnamefont {F.~J.}\ \bibnamefont {Himpsel}},\ }\bibfield  {title}
  {\bibinfo {title} {Fractional band filling in an atomic chain structure},\
  }\href {https://doi.org/10.1103/PhysRevLett.90.176805} {\bibfield  {journal}
  {\bibinfo  {journal} {Phys. Rev. Lett.}\ }\textbf {\bibinfo {volume} {90}},\
  \bibinfo {pages} {176805} (\bibinfo {year} {2003})}\BibitemShut {NoStop}%
\bibitem [{\citenamefont {Yeom}\ \emph {et~al.}(1999)\citenamefont {Yeom},
  \citenamefont {Takeda}, \citenamefont {Rotenberg}, \citenamefont {Matsuda},
  \citenamefont {Horikoshi}, \citenamefont {Schaefer}, \citenamefont {Lee},
  \citenamefont {Kevan}, \citenamefont {Ohta}, \citenamefont {Nagao},\ and\
  \citenamefont {Hasegawa}}]{Yeom1999}%
  \BibitemOpen
  \bibfield  {author} {\bibinfo {author} {\bibfnamefont {H.~W.}\ \bibnamefont
  {Yeom}}, \bibinfo {author} {\bibfnamefont {S.}~\bibnamefont {Takeda}},
  \bibinfo {author} {\bibfnamefont {E.}~\bibnamefont {Rotenberg}}, \bibinfo
  {author} {\bibfnamefont {I.}~\bibnamefont {Matsuda}}, \bibinfo {author}
  {\bibfnamefont {K.}~\bibnamefont {Horikoshi}}, \bibinfo {author}
  {\bibfnamefont {J.}~\bibnamefont {Schaefer}}, \bibinfo {author}
  {\bibfnamefont {C.~M.}\ \bibnamefont {Lee}}, \bibinfo {author} {\bibfnamefont
  {S.~D.}\ \bibnamefont {Kevan}}, \bibinfo {author} {\bibfnamefont
  {T.}~\bibnamefont {Ohta}}, \bibinfo {author} {\bibfnamefont {T.}~\bibnamefont
  {Nagao}},\ and\ \bibinfo {author} {\bibfnamefont {S.}~\bibnamefont
  {Hasegawa}},\ }\bibfield  {title} {\bibinfo {title} {Instability and charge
  density wave of metallic quantum chains on a silicon surface},\ }\href
  {https://doi.org/10.1103/PhysRevLett.82.4898} {\bibfield  {journal} {\bibinfo
   {journal} {Phys. Rev. Lett.}\ }\textbf {\bibinfo {volume} {82}},\ \bibinfo
  {pages} {4898} (\bibinfo {year} {1999})}\BibitemShut {NoStop}%
\bibitem [{\citenamefont {Ahn}\ \emph {et~al.}(2004)\citenamefont {Ahn},
  \citenamefont {Byun}, \citenamefont {Koh}, \citenamefont {Rotenberg},
  \citenamefont {Kevan},\ and\ \citenamefont {Yeom}}]{Ahn2004}%
  \BibitemOpen
  \bibfield  {author} {\bibinfo {author} {\bibfnamefont {J.~R.}\ \bibnamefont
  {Ahn}}, \bibinfo {author} {\bibfnamefont {J.~H.}\ \bibnamefont {Byun}},
  \bibinfo {author} {\bibfnamefont {H.}~\bibnamefont {Koh}}, \bibinfo {author}
  {\bibfnamefont {E.}~\bibnamefont {Rotenberg}}, \bibinfo {author}
  {\bibfnamefont {S.~D.}\ \bibnamefont {Kevan}},\ and\ \bibinfo {author}
  {\bibfnamefont {H.~W.}\ \bibnamefont {Yeom}},\ }\bibfield  {title} {\bibinfo
  {title} {Mechanism of gap opening in a triple-band {Peierls} system: In
  atomic wires on {Si}},\ }\href
  {https://doi.org/10.1103/PhysRevLett.93.106401} {\bibfield  {journal}
  {\bibinfo  {journal} {Phys. Rev. Lett.}\ }\textbf {\bibinfo {volume} {93}},\
  \bibinfo {pages} {106401} (\bibinfo {year} {2004})}\BibitemShut {NoStop}%
\bibitem [{\citenamefont {Tegenkamp}\ \emph {et~al.}(2008)\citenamefont
  {Tegenkamp}, \citenamefont {Ohta}, \citenamefont {McChesney}, \citenamefont
  {Dil}, \citenamefont {Rotenberg}, \citenamefont {Pfn\"ur},\ and\
  \citenamefont {Horn}}]{Tegenkamp2008}%
  \BibitemOpen
  \bibfield  {author} {\bibinfo {author} {\bibfnamefont {C.}~\bibnamefont
  {Tegenkamp}}, \bibinfo {author} {\bibfnamefont {T.}~\bibnamefont {Ohta}},
  \bibinfo {author} {\bibfnamefont {J.~L.}\ \bibnamefont {McChesney}}, \bibinfo
  {author} {\bibfnamefont {H.}~\bibnamefont {Dil}}, \bibinfo {author}
  {\bibfnamefont {E.}~\bibnamefont {Rotenberg}}, \bibinfo {author}
  {\bibfnamefont {H.}~\bibnamefont {Pfn\"ur}},\ and\ \bibinfo {author}
  {\bibfnamefont {K.}~\bibnamefont {Horn}},\ }\bibfield  {title} {\bibinfo
  {title} {Coupled {Pb} chains on {Si(557)}: Origin of one-dimensional
  conductance},\ }\href {https://doi.org/10.1103/PhysRevLett.100.076802}
  {\bibfield  {journal} {\bibinfo  {journal} {Phys. Rev. Lett.}\ }\textbf
  {\bibinfo {volume} {100}},\ \bibinfo {pages} {076802} (\bibinfo {year}
  {2008})}\BibitemShut {NoStop}%
\bibitem [{\citenamefont {Kim}\ \emph {et~al.}(2007)\citenamefont {Kim},
  \citenamefont {Morikawa}, \citenamefont {Choi},\ and\ \citenamefont
  {Yeom}}]{Kim2007}%
  \BibitemOpen
  \bibfield  {author} {\bibinfo {author} {\bibfnamefont {K.~S.}\ \bibnamefont
  {Kim}}, \bibinfo {author} {\bibfnamefont {H.}~\bibnamefont {Morikawa}},
  \bibinfo {author} {\bibfnamefont {W.~H.}\ \bibnamefont {Choi}},\ and\
  \bibinfo {author} {\bibfnamefont {H.~W.}\ \bibnamefont {Yeom}},\ }\bibfield
  {title} {\bibinfo {title} {Strong lateral electron coupling of {Pb} nanowires
  on stepped {Si(111)}: Angle-resolved photoemission studies},\ }\href
  {https://doi.org/10.1103/PhysRevLett.99.196804} {\bibfield  {journal}
  {\bibinfo  {journal} {Phys. Rev. Lett.}\ }\textbf {\bibinfo {volume} {99}},\
  \bibinfo {pages} {196804} (\bibinfo {year} {2007})}\BibitemShut {NoStop}%
\bibitem [{\citenamefont {Appelfeller}\ \emph {et~al.}(2023)\citenamefont
  {Appelfeller}, \citenamefont {Franz}, \citenamefont {Karadag}, \citenamefont
  {Kubicki}, \citenamefont {Zielinski}, \citenamefont {Krivenkov},
  \citenamefont {Varykhalov}, \citenamefont {Preobrajenski},\ and\
  \citenamefont {D{\"{a}}hne}}]{Appelfeller2023}%
  \BibitemOpen
  \bibfield  {author} {\bibinfo {author} {\bibfnamefont {S.}~\bibnamefont
  {Appelfeller}}, \bibinfo {author} {\bibfnamefont {M.}~\bibnamefont {Franz}},
  \bibinfo {author} {\bibfnamefont {M.}~\bibnamefont {Karadag}}, \bibinfo
  {author} {\bibfnamefont {M.}~\bibnamefont {Kubicki}}, \bibinfo {author}
  {\bibfnamefont {R.}~\bibnamefont {Zielinski}}, \bibinfo {author}
  {\bibfnamefont {M.}~\bibnamefont {Krivenkov}}, \bibinfo {author}
  {\bibfnamefont {A.}~\bibnamefont {Varykhalov}}, \bibinfo {author}
  {\bibfnamefont {A.}~\bibnamefont {Preobrajenski}},\ and\ \bibinfo {author}
  {\bibfnamefont {M.}~\bibnamefont {D{\"{a}}hne}},\ }\bibfield  {title}
  {\bibinfo {title} {Self-organized formation of unidirectional and
  quasi-one-dimensional metallic {Tb} silicide nanowires on {Si(110)}},\ }\href
  {https://doi.org/https://doi.org/10.1016/j.apsusc.2022.154875} {\bibfield
  {journal} {\bibinfo  {journal} {Appl. Surf. Sci.}\ }\textbf {\bibinfo
  {volume} {607}},\ \bibinfo {pages} {154875} (\bibinfo {year}
  {2023})}\BibitemShut {NoStop}%
\bibitem [{\citenamefont {Pampuch}\ \emph {et~al.}(2000)\citenamefont
  {Pampuch}, \citenamefont {Rader}, \citenamefont {Kachel}, \citenamefont
  {Gudat}, \citenamefont {Carbone}, \citenamefont {Kl\"asges}, \citenamefont
  {Bihlmayer}, \citenamefont {Bl\"ugel},\ and\ \citenamefont
  {Eberhardt}}]{Pampuch2000}%
  \BibitemOpen
  \bibfield  {author} {\bibinfo {author} {\bibfnamefont {C.}~\bibnamefont
  {Pampuch}}, \bibinfo {author} {\bibfnamefont {O.}~\bibnamefont {Rader}},
  \bibinfo {author} {\bibfnamefont {T.}~\bibnamefont {Kachel}}, \bibinfo
  {author} {\bibfnamefont {W.}~\bibnamefont {Gudat}}, \bibinfo {author}
  {\bibfnamefont {C.}~\bibnamefont {Carbone}}, \bibinfo {author} {\bibfnamefont
  {R.}~\bibnamefont {Kl\"asges}}, \bibinfo {author} {\bibfnamefont
  {G.}~\bibnamefont {Bihlmayer}}, \bibinfo {author} {\bibfnamefont
  {S.}~\bibnamefont {Bl\"ugel}},\ and\ \bibinfo {author} {\bibfnamefont
  {W.}~\bibnamefont {Eberhardt}},\ }\bibfield  {title} {\bibinfo {title}
  {One-dimensional spin-polarized quantum-wire states in {Au} on {Ni(110)}},\
  }\href {https://doi.org/10.1103/PhysRevLett.85.2561} {\bibfield  {journal}
  {\bibinfo  {journal} {Phys. Rev. Lett.}\ }\textbf {\bibinfo {volume} {85}},\
  \bibinfo {pages} {2561} (\bibinfo {year} {2000})}\BibitemShut {NoStop}%
\bibitem [{\citenamefont {Bianchi}\ \emph {et~al.}(2015)\citenamefont
  {Bianchi}, \citenamefont {Song}, \citenamefont {Cooil}, \citenamefont
  {Monsen}, \citenamefont {Wahlstr\"om}, \citenamefont {Miwa}, \citenamefont
  {Rienks}, \citenamefont {Evans}, \citenamefont {Strozecka}, \citenamefont
  {Pascual}, \citenamefont {Leandersson}, \citenamefont {Balasubramanian},
  \citenamefont {Hofmann},\ and\ \citenamefont {Wells}}]{Bianchi2015}%
  \BibitemOpen
  \bibfield  {author} {\bibinfo {author} {\bibfnamefont {M.}~\bibnamefont
  {Bianchi}}, \bibinfo {author} {\bibfnamefont {F.}~\bibnamefont {Song}},
  \bibinfo {author} {\bibfnamefont {S.}~\bibnamefont {Cooil}}, \bibinfo
  {author} {\bibfnamefont {A.~F.}\ \bibnamefont {Monsen}}, \bibinfo {author}
  {\bibfnamefont {E.}~\bibnamefont {Wahlstr\"om}}, \bibinfo {author}
  {\bibfnamefont {J.~A.}\ \bibnamefont {Miwa}}, \bibinfo {author}
  {\bibfnamefont {E.~D.~L.}\ \bibnamefont {Rienks}}, \bibinfo {author}
  {\bibfnamefont {D.~A.}\ \bibnamefont {Evans}}, \bibinfo {author}
  {\bibfnamefont {A.}~\bibnamefont {Strozecka}}, \bibinfo {author}
  {\bibfnamefont {J.~I.}\ \bibnamefont {Pascual}}, \bibinfo {author}
  {\bibfnamefont {M.}~\bibnamefont {Leandersson}}, \bibinfo {author}
  {\bibfnamefont {T.}~\bibnamefont {Balasubramanian}}, \bibinfo {author}
  {\bibfnamefont {P.}~\bibnamefont {Hofmann}},\ and\ \bibinfo {author}
  {\bibfnamefont {J.~W.}\ \bibnamefont {Wells}},\ }\bibfield  {title} {\bibinfo
  {title} {One-dimensional spin texture of {Bi(441)}: Quantum spin hall
  properties without a topological insulator},\ }\href
  {https://doi.org/10.1103/PhysRevB.91.165307} {\bibfield  {journal} {\bibinfo
  {journal} {Phys. Rev. B}\ }\textbf {\bibinfo {volume} {91}},\ \bibinfo
  {pages} {165307} (\bibinfo {year} {2015})}\BibitemShut {NoStop}%
\bibitem [{\citenamefont {Grothe}\ \emph {et~al.}(2013)\citenamefont {Grothe},
  \citenamefont {Johnston}, \citenamefont {Chi}, \citenamefont {Dosanjh},
  \citenamefont {Burke},\ and\ \citenamefont {Pennec}}]{AgSS_Grothe2013}%
  \BibitemOpen
  \bibfield  {author} {\bibinfo {author} {\bibfnamefont {S.}~\bibnamefont
  {Grothe}}, \bibinfo {author} {\bibfnamefont {S.}~\bibnamefont {Johnston}},
  \bibinfo {author} {\bibfnamefont {S.}~\bibnamefont {Chi}}, \bibinfo {author}
  {\bibfnamefont {P.}~\bibnamefont {Dosanjh}}, \bibinfo {author} {\bibfnamefont
  {S.~A.}\ \bibnamefont {Burke}},\ and\ \bibinfo {author} {\bibfnamefont
  {Y.}~\bibnamefont {Pennec}},\ }\bibfield  {title} {\bibinfo {title}
  {Quantifying many-body effects by high-resolution {Fourier} transform
  scanning tunneling spectroscopy},\ }\href
  {https://doi.org/10.1103/PhysRevLett.111.246804} {\bibfield  {journal}
  {\bibinfo  {journal} {Phys. Rev. Lett.}\ }\textbf {\bibinfo {volume} {111}},\
  \bibinfo {pages} {246804} (\bibinfo {year} {2013})}\BibitemShut {NoStop}%
\bibitem [{Sup()}]{Supplement}%
  \BibitemOpen
  \href@noop {} {\bibinfo {title} {Supplementary information is available
  online.}}\BibitemShut {Stop}%
\bibitem [{\citenamefont {H\"{o}vel}\ \emph {et~al.}(2001)\citenamefont
  {H\"{o}vel}, \citenamefont {Grimm},\ and\ \citenamefont
  {Reihl}}]{STS-QIP-SS}%
  \BibitemOpen
  \bibfield  {author} {\bibinfo {author} {\bibfnamefont {H.}~\bibnamefont
  {H\"{o}vel}}, \bibinfo {author} {\bibfnamefont {B.}~\bibnamefont {Grimm}},\
  and\ \bibinfo {author} {\bibfnamefont {B.}~\bibnamefont {Reihl}},\ }\bibfield
   {title} {\bibinfo {title} {Modification of the {Shockley}-type surface state
  on {Ag(111)} by an adsorbed xenon layer},\ }\href
  {https://doi.org/10.1016/s0039-6028(01)00704-x} {\bibfield  {journal}
  {\bibinfo  {journal} {Surf. Sci.}\ }\textbf {\bibinfo {volume} {477}},\
  \bibinfo {pages} {43} (\bibinfo {year} {2001})}\BibitemShut {NoStop}%
\bibitem [{\citenamefont {Krivenkov}\ \emph {et~al.}(2022)\citenamefont
  {Krivenkov}, \citenamefont {Marchenko}, \citenamefont {Sajedi}, \citenamefont
  {Fedorov}, \citenamefont {Clark}, \citenamefont {Sánchez-Barriga},
  \citenamefont {Rienks}, \citenamefont {Rader},\ and\ \citenamefont
  {Varykhalov}}]{Krivenkov-C60-Au111}%
  \BibitemOpen
  \bibfield  {author} {\bibinfo {author} {\bibfnamefont {M.}~\bibnamefont
  {Krivenkov}}, \bibinfo {author} {\bibfnamefont {D.}~\bibnamefont
  {Marchenko}}, \bibinfo {author} {\bibfnamefont {M.}~\bibnamefont {Sajedi}},
  \bibinfo {author} {\bibfnamefont {A.}~\bibnamefont {Fedorov}}, \bibinfo
  {author} {\bibfnamefont {O.~J.}\ \bibnamefont {Clark}}, \bibinfo {author}
  {\bibfnamefont {J.}~\bibnamefont {Sánchez-Barriga}}, \bibinfo {author}
  {\bibfnamefont {E.~D.~L.}\ \bibnamefont {Rienks}}, \bibinfo {author}
  {\bibfnamefont {O.}~\bibnamefont {Rader}},\ and\ \bibinfo {author}
  {\bibfnamefont {A.}~\bibnamefont {Varykhalov}},\ }\bibfield  {title}
  {\bibinfo {title} {On the problem of dirac cones in fullerenes on gold},\
  }\href {https://doi.org/10.1039/D1NR07981F} {\bibfield  {journal} {\bibinfo
  {journal} {Nanoscale}\ }\textbf {\bibinfo {volume} {14}},\ \bibinfo {pages}
  {9124} (\bibinfo {year} {2022})}\BibitemShut {NoStop}%
\bibitem [{\citenamefont {Tani}\ \emph {et~al.}(2022)\citenamefont {Tani},
  \citenamefont {Hitomi}, \citenamefont {Kawakami},\ and\ \citenamefont
  {Koshino}}]{BlueP-Tani-PRB-2022}%
  \BibitemOpen
  \bibfield  {author} {\bibinfo {author} {\bibfnamefont {T.}~\bibnamefont
  {Tani}}, \bibinfo {author} {\bibfnamefont {M.}~\bibnamefont {Hitomi}},
  \bibinfo {author} {\bibfnamefont {T.}~\bibnamefont {Kawakami}},\ and\
  \bibinfo {author} {\bibfnamefont {M.}~\bibnamefont {Koshino}},\ }\bibfield
  {title} {\bibinfo {title} {Topological edge and corner states and fractional
  corner charges in blue phosphorene},\ }\href
  {https://doi.org/10.1103/physrevb.105.075407} {\bibfield  {journal} {\bibinfo
   {journal} {Phys. Rev. B}\ }\textbf {\bibinfo {volume} {105}},\ \bibinfo
  {pages} {075407} (\bibinfo {year} {2022})}\BibitemShut {NoStop}%
\end{thebibliography}%


\begin{thebibliography}{13}%
\makeatletter
\providecommand \@ifxundefined [1]{%
 \@ifx{#1\undefined}
}%
\providecommand \@ifnum [1]{%
 \ifnum #1\expandafter \@firstoftwo
 \else \expandafter \@secondoftwo
 \fi
}%
\providecommand \@ifx [1]{%
 \ifx #1\expandafter \@firstoftwo
 \else \expandafter \@secondoftwo
 \fi
}%
\providecommand \natexlab [1]{#1}%
\providecommand \enquote  [1]{``#1''}%
\providecommand \bibnamefont  [1]{#1}%
\providecommand \bibfnamefont [1]{#1}%
\providecommand \citenamefont [1]{#1}%
\providecommand \href@noop [0]{\@secondoftwo}%
\providecommand \href [0]{\begingroup \@sanitize@url \@href}%
\providecommand \@href[1]{\@@startlink{#1}\@@href}%
\providecommand \@@href[1]{\endgroup#1\@@endlink}%
\providecommand \@sanitize@url [0]{\catcode `\\12\catcode `\$12\catcode
  `\&12\catcode `\#12\catcode `\^12\catcode `\_12\catcode `\%12\relax}%
\providecommand \@@startlink[1]{}%
\providecommand \@@endlink[0]{}%
\providecommand \url  [0]{\begingroup\@sanitize@url \@url }%
\providecommand \@url [1]{\endgroup\@href {#1}{\urlprefix }}%
\providecommand \urlprefix  [0]{URL }%
\providecommand \Eprint [0]{\href }%
\providecommand \doibase [0]{https://doi.org/}%
\providecommand \selectlanguage [0]{\@gobble}%
\providecommand \bibinfo  [0]{\@secondoftwo}%
\providecommand \bibfield  [0]{\@secondoftwo}%
\providecommand \translation [1]{[#1]}%
\providecommand \BibitemOpen [0]{}%
\providecommand \bibitemStop [0]{}%
\providecommand \bibitemNoStop [0]{.\EOS\space}%
\providecommand \EOS [0]{\spacefactor3000\relax}%
\providecommand \BibitemShut  [1]{\csname bibitem#1\endcsname}%
\let\auto@bib@innerbib\@empty
\bibitem [{\citenamefont {Zhang}\ \emph {et~al.}(2021)\citenamefont {Zhang},
  \citenamefont {Enriquez}, \citenamefont {Tong}, \citenamefont {Mayne},
  \citenamefont {Bendounan}, \citenamefont {Smogunov}, \citenamefont {Dappe},
  \citenamefont {Kara}, \citenamefont {Dujardin},\ and\ \citenamefont
  {Oughaddou}}]{BlueP-Zhang-NatComm-2021}%
  \BibitemOpen
  \bibfield  {author} {\bibinfo {author} {\bibfnamefont {W.}~\bibnamefont
  {Zhang}}, \bibinfo {author} {\bibfnamefont {H.}~\bibnamefont {Enriquez}},
  \bibinfo {author} {\bibfnamefont {Y.}~\bibnamefont {Tong}}, \bibinfo {author}
  {\bibfnamefont {A.~J.}\ \bibnamefont {Mayne}}, \bibinfo {author}
  {\bibfnamefont {A.}~\bibnamefont {Bendounan}}, \bibinfo {author}
  {\bibfnamefont {A.}~\bibnamefont {Smogunov}}, \bibinfo {author}
  {\bibfnamefont {Y.~J.}\ \bibnamefont {Dappe}}, \bibinfo {author}
  {\bibfnamefont {A.}~\bibnamefont {Kara}}, \bibinfo {author} {\bibfnamefont
  {G.}~\bibnamefont {Dujardin}},\ and\ \bibinfo {author} {\bibfnamefont
  {H.}~\bibnamefont {Oughaddou}},\ }\bibfield  {title} {\bibinfo {title} {Flat
  epitaxial quasi-{1D} phosphorene chains},\ }\href
  {https://doi.org/10.1038/s41467-021-25262-7} {\bibfield  {journal} {\bibinfo
  {journal} {Nat. Commun.}\ }\textbf {\bibinfo {volume} {12}},\ \bibinfo
  {pages} {5160} (\bibinfo {year} {2021})}\BibitemShut {NoStop}%
\bibitem [{\citenamefont {Wilson}(1962)}]{Wilson1962}%
  \BibitemOpen
  \bibfield  {author} {\bibinfo {author} {\bibfnamefont {A.~J.~C.}\
  \bibnamefont {Wilson}},\ }\href@noop {} {\emph {\bibinfo {title} {{X}-ray
  Optics: the Diffraction of {X}-rays by Finite and Imperfect Crystals}}}\
  (\bibinfo  {publisher} {Methuen},\ \bibinfo {year} {1962})\BibitemShut
  {NoStop}%
\bibitem [{\citenamefont {Schroeder}\ \emph {et~al.}(2002)\citenamefont
  {Schroeder}, \citenamefont {Giorgi}, \citenamefont {Hammoudeh}, \citenamefont
  {Magg}, \citenamefont {B{\"{a}}umer},\ and\ \citenamefont
  {Freund}}]{Schroeder2002}%
  \BibitemOpen
  \bibfield  {author} {\bibinfo {author} {\bibfnamefont {T.}~\bibnamefont
  {Schroeder}}, \bibinfo {author} {\bibfnamefont {J.~B.}\ \bibnamefont
  {Giorgi}}, \bibinfo {author} {\bibfnamefont {A.}~\bibnamefont {Hammoudeh}},
  \bibinfo {author} {\bibfnamefont {N.}~\bibnamefont {Magg}}, \bibinfo {author}
  {\bibfnamefont {M.}~\bibnamefont {B{\"{a}}umer}},\ and\ \bibinfo {author}
  {\bibfnamefont {H.-J.}\ \bibnamefont {Freund}},\ }\bibfield  {title}
  {\bibinfo {title} {Oxygen-induced p$(2\times3)$ reconstruction on {Mo}(112)
  studied by {LEED} and {STM}},\ }\href
  {https://doi.org/10.1103/PhysRevB.65.115411} {\bibfield  {journal} {\bibinfo
  {journal} {Phys. Rev. B}\ }\textbf {\bibinfo {volume} {65}},\ \bibinfo
  {pages} {115411} (\bibinfo {year} {2002})}\BibitemShut {NoStop}%
\bibitem [{\citenamefont {Hafke}\ \emph {et~al.}(2016)\citenamefont {Hafke},
  \citenamefont {Frigge}, \citenamefont {Witte}, \citenamefont {Krenzer},
  \citenamefont {Aulbach}, \citenamefont {Sch{\"{a}}fer}, \citenamefont
  {Claessen}, \citenamefont {Erwin},\ and\ \citenamefont {Horn-von
  Hoegen}}]{Hafke2016}%
  \BibitemOpen
  \bibfield  {author} {\bibinfo {author} {\bibfnamefont {B.}~\bibnamefont
  {Hafke}}, \bibinfo {author} {\bibfnamefont {T.}~\bibnamefont {Frigge}},
  \bibinfo {author} {\bibfnamefont {T.}~\bibnamefont {Witte}}, \bibinfo
  {author} {\bibfnamefont {B.}~\bibnamefont {Krenzer}}, \bibinfo {author}
  {\bibfnamefont {J.}~\bibnamefont {Aulbach}}, \bibinfo {author} {\bibfnamefont
  {J.}~\bibnamefont {Sch{\"{a}}fer}}, \bibinfo {author} {\bibfnamefont
  {R.}~\bibnamefont {Claessen}}, \bibinfo {author} {\bibfnamefont {S.~C.}\
  \bibnamefont {Erwin}},\ and\ \bibinfo {author} {\bibfnamefont
  {M.}~\bibnamefont {Horn-von Hoegen}},\ }\bibfield  {title} {\bibinfo {title}
  {Two-dimensional interaction of spin chains in the {Si}(553)-{Au} nanowire
  system},\ }\href {https://doi.org/10.1103/PhysRevB.94.161403} {\bibfield
  {journal} {\bibinfo  {journal} {Phys. Rev. B}\ }\textbf {\bibinfo {volume}
  {94}},\ \bibinfo {pages} {161403} (\bibinfo {year} {2016})}\BibitemShut
  {NoStop}%
\bibitem [{ARP()}]{ARPES-robotics}%
  \BibitemOpen
  \href {www.arpes-robotics.com} {\bibinfo {title}
  {www.arpes-robotics.com}}\BibitemShut {NoStop}%
\bibitem [{\citenamefont {Varykhalov}\ \emph {et~al.}(2005)\citenamefont
  {Varykhalov}, \citenamefont {Rader},\ and\ \citenamefont
  {Gudat}}]{Varykhalov-PRB-2005}%
  \BibitemOpen
  \bibfield  {author} {\bibinfo {author} {\bibfnamefont {A.}~\bibnamefont
  {Varykhalov}}, \bibinfo {author} {\bibfnamefont {O.}~\bibnamefont {Rader}},\
  and\ \bibinfo {author} {\bibfnamefont {W.}~\bibnamefont {Gudat}},\ }\bibfield
   {title} {\bibinfo {title} {Structure and quantum-size effects in a surface
  carbide:
  $\mathrm{W}(110)/\mathrm{C}\text{\ensuremath{-}}{R}(15\ifmmode\times\else\texttimes\fi{}3)$},\
  }\href {https://doi.org/10.1103/PhysRevB.72.115440} {\bibfield  {journal}
  {\bibinfo  {journal} {Phys. Rev. B}\ }\textbf {\bibinfo {volume} {72}},\
  \bibinfo {pages} {115440} (\bibinfo {year} {2005})}\BibitemShut {NoStop}%
\bibitem [{\citenamefont {Kresse}\ and\ \citenamefont {Hafner}(1993)}]{Kresse}%
  \BibitemOpen
  \bibfield  {author} {\bibinfo {author} {\bibfnamefont {G.}~\bibnamefont
  {Kresse}}\ and\ \bibinfo {author} {\bibfnamefont {J.}~\bibnamefont
  {Hafner}},\ }\bibfield  {title} {\bibinfo {title} {Ab initio molecular
  dynamics for liquid metals},\ }\href
  {https://doi.org/10.1103/physrevb.47.558} {\bibfield  {journal} {\bibinfo
  {journal} {Phys. Rev. B}\ }\textbf {\bibinfo {volume} {47}},\ \bibinfo
  {pages} {558} (\bibinfo {year} {1993})}\BibitemShut {NoStop}%
\bibitem [{\citenamefont {Grimme}(2006)}]{Grimme}%
  \BibitemOpen
  \bibfield  {author} {\bibinfo {author} {\bibfnamefont {S.}~\bibnamefont
  {Grimme}},\ }\bibfield  {title} {\bibinfo {title} {Semiempirical {GGA}-type
  density functional constructed with a long-range dispersion correction},\
  }\href {https://doi.org/10.1002/jcc.20495} {\bibfield  {journal} {\bibinfo
  {journal} {J. Comput. Chem.}\ }\textbf {\bibinfo {volume} {27}},\ \bibinfo
  {pages} {1787} (\bibinfo {year} {2006})}\BibitemShut {NoStop}%
\bibitem [{Unf()}]{Unfolding}%
  \BibitemOpen
  \href {https://github.com/QijingZheng/VaspBandUnfolding} {\bibinfo {title}
  {{https://github.com/QijingZheng/VaspBandUnfolding}}}\BibitemShut {NoStop}%
\bibitem [{\citenamefont {Momma}\ and\ \citenamefont {Izumi}(2011)}]{VESTA}%
  \BibitemOpen
  \bibfield  {author} {\bibinfo {author} {\bibfnamefont {K.}~\bibnamefont
  {Momma}}\ and\ \bibinfo {author} {\bibfnamefont {F.}~\bibnamefont {Izumi}},\
  }\bibfield  {title} {\bibinfo {title} {{\it VESTA3} for three-dimensional
  visualization of crystal, volumetric and morphology data},\ }\href
  {https://doi.org/10.1107/S0021889811038970} {\bibfield  {journal} {\bibinfo
  {journal} {Journal of Applied Crystallography}\ }\textbf {\bibinfo {volume}
  {44}},\ \bibinfo {pages} {1272} (\bibinfo {year} {2011})}\BibitemShut
  {NoStop}%
\bibitem [{\citenamefont {Ne{\v{c}}as}\ and\ \citenamefont
  {Klapetek}(2012)}]{Necas2012}%
  \BibitemOpen
  \bibfield  {author} {\bibinfo {author} {\bibfnamefont {D.}~\bibnamefont
  {Ne{\v{c}}as}}\ and\ \bibinfo {author} {\bibfnamefont {P.}~\bibnamefont
  {Klapetek}},\ }\bibfield  {title} {\bibinfo {title} {Gwyddion: an open-source
  software for {SPM} data analysis},\ }\href
  {https://doi.org/10.2478/s11534-011-0096-2} {\bibfield  {journal} {\bibinfo
  {journal} {Central European Journal of Physics}\ }\textbf {\bibinfo {volume}
  {10}},\ \bibinfo {pages} {181} (\bibinfo {year} {2012})}\BibitemShut
  {NoStop}%
\bibitem [{Gwy()}]{Gwyddion}%
  \BibitemOpen
  \href@noop {} {\bibinfo {title}
  {\href{http://gwyddion.net/}{http://gwyddion.net/}}}\BibitemShut {NoStop}%
\bibitem [{LEE(2014)}]{LEEDpat42}%
  \BibitemOpen
  \href@noop {} {\bibinfo {title} {{LEEDpat, Version 4.2, utility by K.E.
  Hermann (FHI) and M.A. Van Hove (HKBU), Berlin / Hong Kong}; 2014.
  \href{http://www.fhi-berlin.mpg.de/KHsoftware/LEEDpat/index.html}{http://www.fhi-berlin.mpg.de/KHsoftware/LEEDpat/index.html}}}
  (\bibinfo {year} {2014})\BibitemShut {NoStop}%
\end{thebibliography}%


\end{document}


\def\blue{\textcolor{blue}}
	\def\gr{\textcolor[rgb]{0,0.6,0}}
	\def\br{\textcolor[rgb]{0.7,0.5,0}}
	\def\prp{\textcolor[rgb]{0.5,0,0.5}}

	\def\LDP{LD-phase}
	\def\HDP{HD-phase}
	\def\AB{{\it A-B}}
	\def\Ef{$E_{\rm F}$}
	\def\Tc{$T_{\rm C}$}
	\def\kpara{{\bf k}$_\parallel$}
	\def\kparax{{\bf k}$_{\parallel,x}$}
	\def\kparay{{\bf k}$_{\parallel,y}$}
	\def\kperp{{\bf k}$_\perp$}
	\def\dirGX{$\overline{\rm \Gamma}-\overline{\rm X}$}
	\def\dirGY{$\overline{\rm \Gamma}-\overline{\rm Y}$}
	\def\dirGK{$\overline{\rm \Gamma}-\overline{\rm K}_{Ag}$}
	\def\dirGKprime{$\overline{\rm \Gamma}-\overline{\rm K}_{Ag}'$}
	\def\dirGM{$\overline{\rm \Gamma}-\overline{\rm M}_{Ag}$}
	\def\dirGS{$\overline{\rm \Gamma}-\overline{\rm S}$}
	\def\dirGN{$\overline{\rm \Gamma}-\overline{\rm N}$}
	\def\dirGNhalf{$\frac{1}{2}(\overline{\rm \Gamma}-\overline{\rm N})$}
	\def\pntG{$\overline{\rm \Gamma}$}
	\def\pntM{$\overline{\rm M}_{Ag}$}
	\def\pntMprime{$\overline{\rm M'}$}
	\def\pntK{$\overline{\rm K}_{Ag}$}
	\def\pntN{$\overline{\rm N}$}
	\def\pntNhalf{ $\overline{\rm N}/2$ }
	\def\invA{\AA$^{-1}$}
	\def\DCgamma{${\rm DC}_{\overline{\Gamma}}$}
	\def\DCNhalf{${\rm DC}_{\overline{\rm N}/{\rm 2}}$}
	
	\def\root33{$\sqrt{3}\times\sqrt{3}$ {\it R}30$^\circ$}
	\def\RT3{$\sqrt{3}$}

	\renewcommand{\andname}{\ignorespaces}
	
	\title{\textit{Supporting information for}\\Truly one-dimensional electronic structure of phosphorene chains}

	\author{Maxim Krivenkov$^{1,\dagger}$}\email[Corresponding author: ]{maxim.krivenkov@helmholtz-berlin.de}
	\author{Maryam Sajedi$^{1,\dagger}$}
	\author{Dmitry Marchenko$^1$}
	\author{Evangelos Golias$^2$}
	\author{Matthias Muntwiler$^3$}
	\author{Oliver Rader$^{1}$ and Andrei Varykhalov$^{1}$}
	
	\affiliation{$^1$ Helmholtz-Zentrum Berlin f\"ur Materialien und Energie,Elektronenspeicherring BESSY II, Albert-Einstein-Str. 15, 12489 Berlin, Germany}
	\affiliation{$^2$ MAX IV Laboratory, Lund University, Fotongatan 2, 22484, Lund, Sweden}
	\affiliation{$^3$ Paul Scherrer Institute, 5232 Villigen, Switzerland}
	\affiliation{$^\dagger$ These authors contributed equally to this work}
	
	\maketitle

	\begin{figure*}[t]
		\centering
		\includegraphics[width=0.95\textwidth]{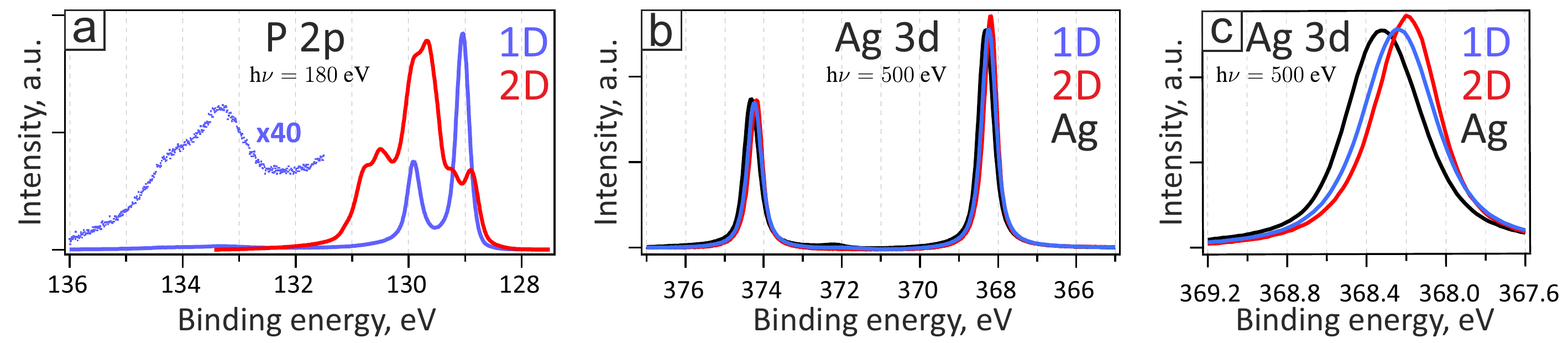}
		\caption{(a) {\it P 2p} and (b,c) {\it Ag 3d} core levels for P chains (blue) compared to 2D phosphorene on Ag(111) (red) and clean Ag(111) (black). (c) Zoom-in of the right peak in (b) showing minor chemical shifts. For all spectra a Shirley background was subtracted. We observe additional weak doublets with splitting similar to P 2p and Ag 3d, but shifted by $\sim$4 eV to higher binding energies.
		}
	\end{figure*}
	
	\begin{figure*}[t]
		\centering
		\includegraphics[width=0.90\textwidth]{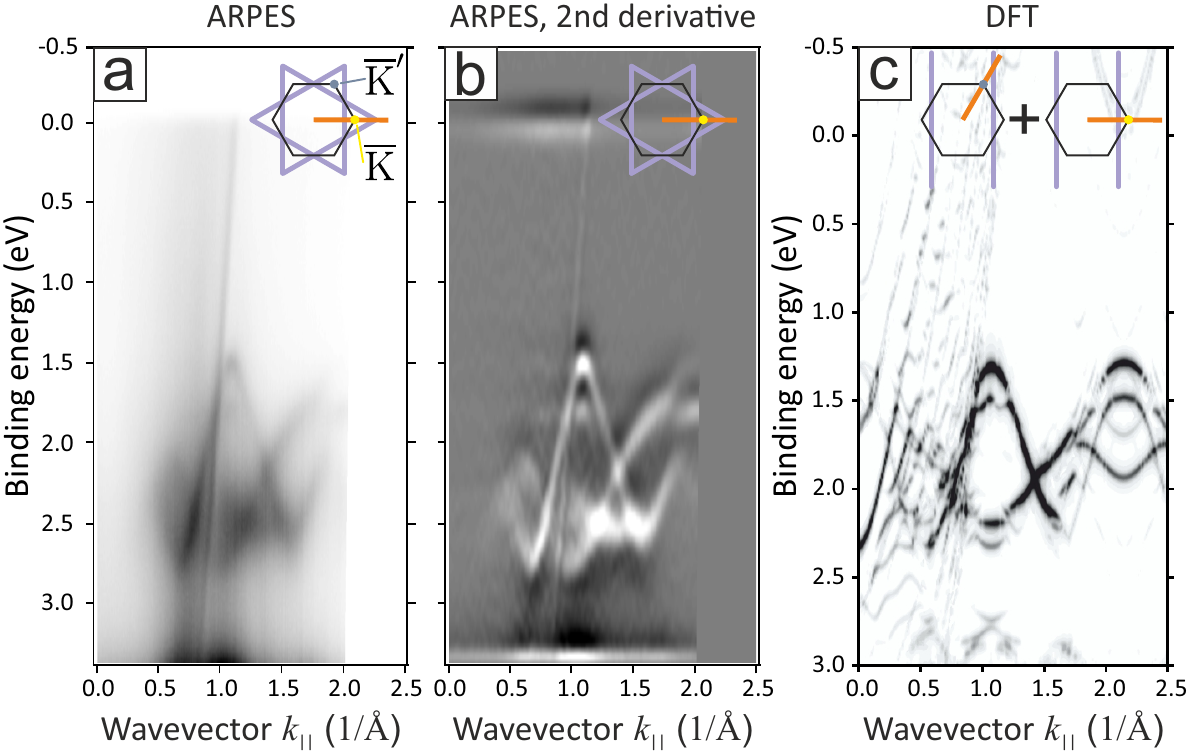}
		\caption{Comparison of ARPES data with the band structures calculated in different directions and overlapped to simulate the effect of domains. (a--b) ARPES measurements along \dirGK\ direction, raw intensity (a) and second derivative (b). (c) Unfolded band structure calculations along the \dirGK\ direction which is along the wire overlapped with the unfolded calculations along \dirGKprime\ direction, which cuts the P bands at an angle of 60$^\circ$. 
		The second direction corresponds to the signal from rotated domains. The projection only on P atoms is shown. The inset demonstrates the SBZ of Ag(111) (black), the position of P valence band maxima for three different domains (blue) and the measurement or calculation direction (orange).
		}
	\end{figure*}

	\begin{figure*}[t]
		\centering
		\includegraphics[width=0.82\textwidth]{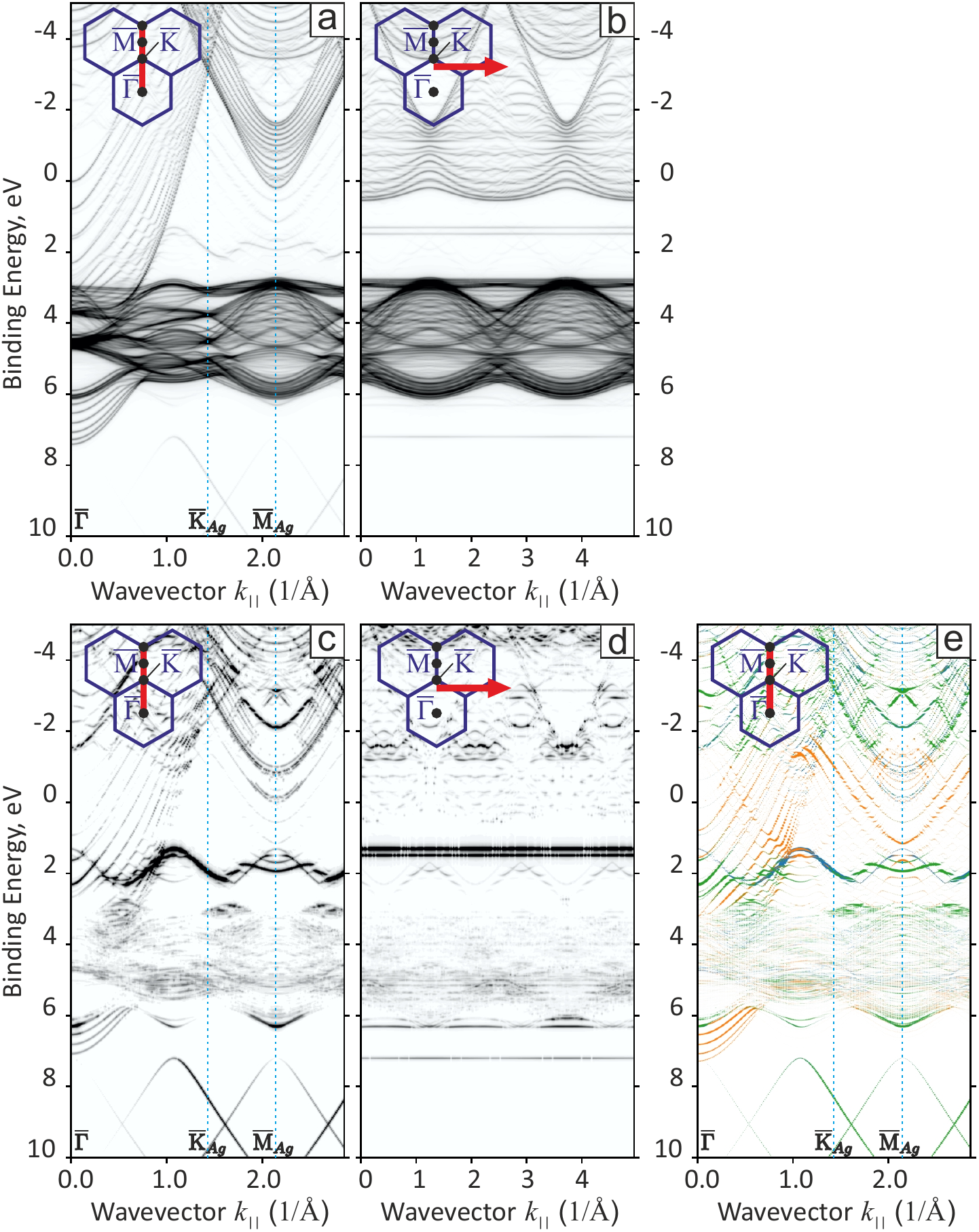}
		\caption{
			Unfolded DFT band structure calculations of P chains on Ag(111) similar to Fig. 4 of the main text, but in wider energy range. 
			(a,b) Combined bands of P and Ag; (c,d) Projection of the band structure on phosphorus atoms only.
			Along \dirGK\ direction of Ag(111) which is parallel to the chains (a,c) and along the perpendicular direction passing through the valence band maximum (b,d), the directions are marked in the insets by red lines. Panel (e) is similar to (c), but size and color of markers demonstrates projections onto $p_x$ (green), $p_y$ (blue) and $p_z$ (orange) orbitals, where the $x$ axis is along the chains. The P band visible at 7--10 eV binding energy actually has mainly $s$ character (not shown). 
		}
	\end{figure*}
	
	\begin{figure*}[t]
		\centering
		\includegraphics[width=0.95\textwidth]{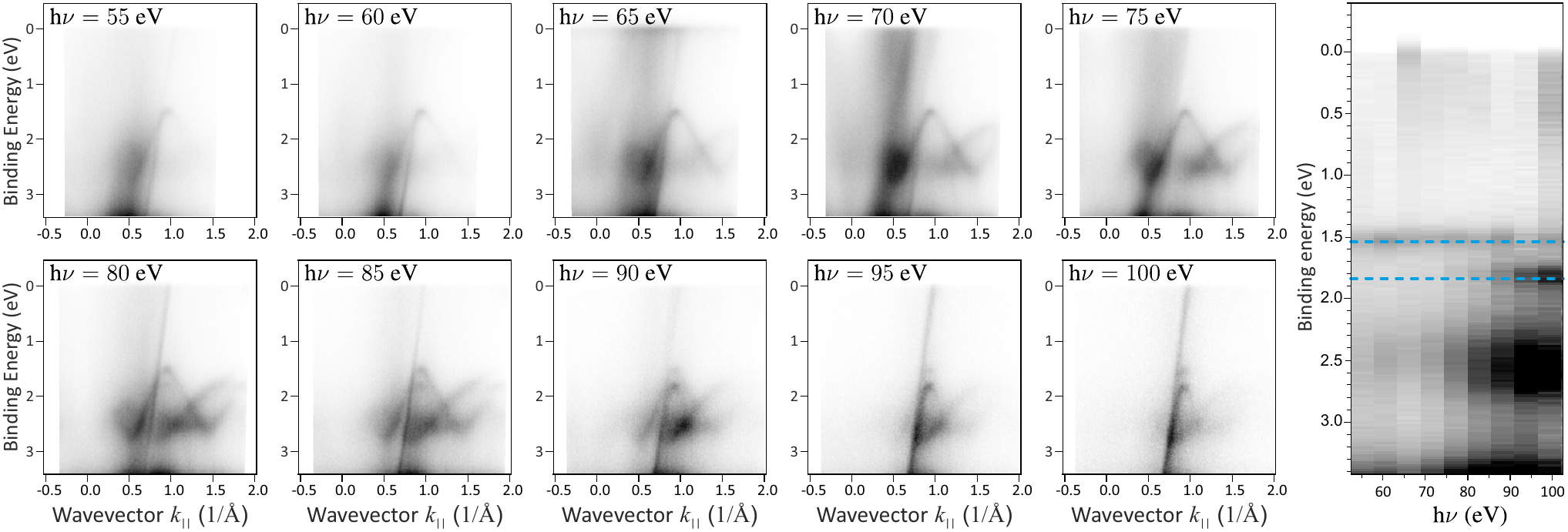}
		\caption{Photon energy dependence of P states. On the left side ARPES spectra measured with photon energies h$\nu$ from 55 eV to 100 eV are presented and on the right a combined plot of energy distribution curves is presented taken at $k=0.94$ \invA\ (at valence band maximum) as a function of h$\nu$. One can see that there is also no dispersion of P bands in the $k_z$ direction (blue dashed lines are guides to the eye).
		}
	\end{figure*}
	
	\begin{figure*}[t]
		\centering
		\includegraphics[width=0.95\textwidth]{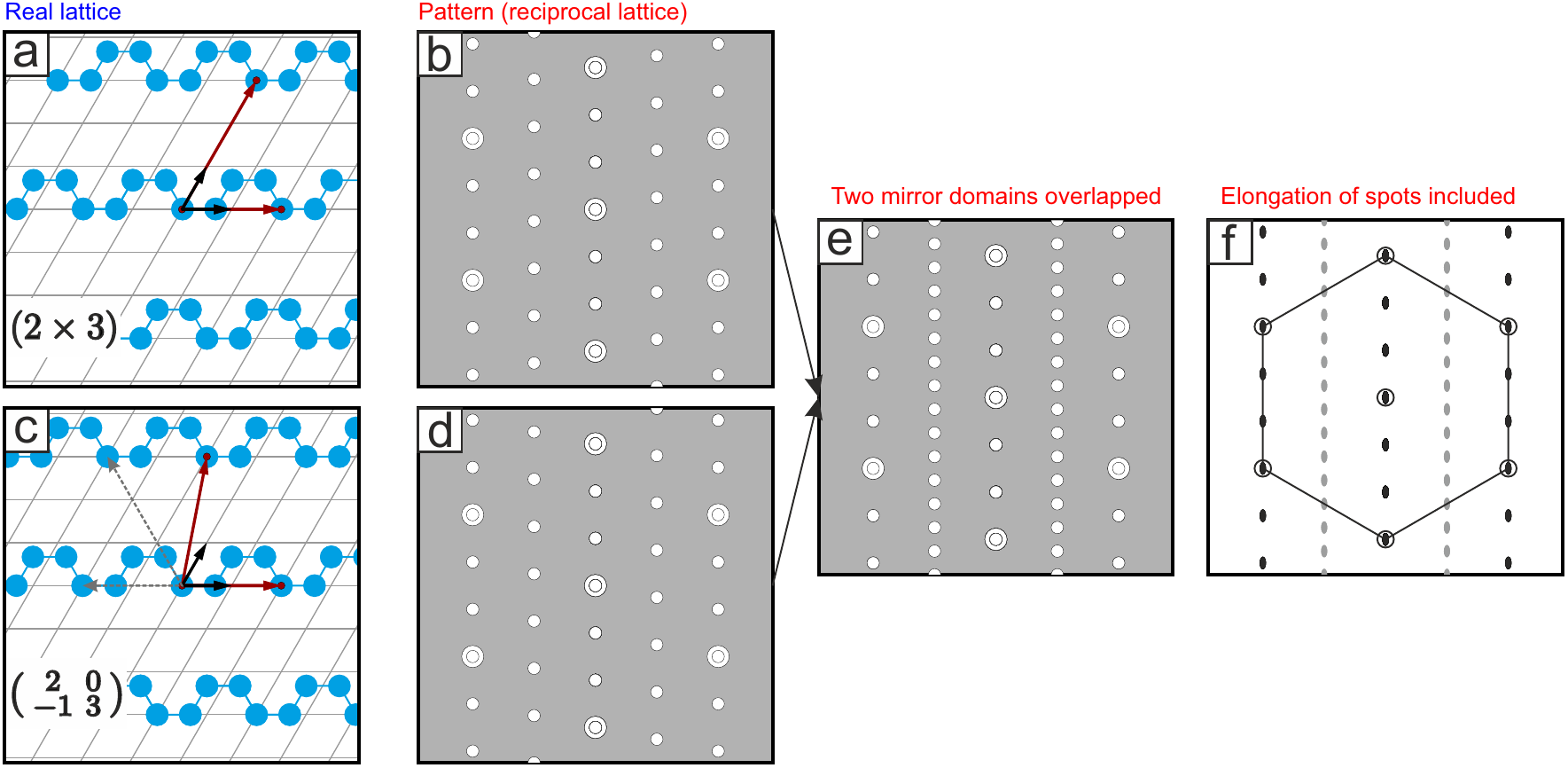}
		\caption{Simplistic model of lines in LEED of P chains. (a,c) Unit cells of $(2\times3) = \smqty(2 & 0 \\ 0 & 3)$ and $\smqty(2 & 0 \\ -1 & 3)$ superlattices and their reciprocal lattices (b,d), respectively (large spots correspond to Ag(111) substrate). When these two reciprocal lattices are overlapped (e), dense rows of spots appear at $\times2$ position and when spots are elongated as in the experiment (f), they nearly join together and resemble lines.
		}
		\label{fig:LEED}
	\end{figure*}
	
	\clearpage
	The main idea that helps to explain the LEED pattern of P chains is presented in Fig.~\ref{fig:LEED}. There are actually two slightly different structures of P chains \cite{BlueP-Zhang-NatComm-2021}: if the first one is a simple $(2\times3)$ [Fig.~\ref{fig:LEED}(a)], in the second one the chains are shifted relative to each other along the wire by 1 lattice period of Ag(111) or 1/2 of the chain supercell [Fig.~\ref{fig:LEED}(c)]. These two P structures are related to each other by mirroring along $x$ axis [grey dashed vectors in (c) are mirrored supercell vectors from (a)]. The corresponding reciprocal lattices [Fig.~\ref{fig:LEED}(b,d)] are also mirror images of each other and are \textit{not identical}. In the LEED experiment the electron beam spot is sufficiently large and covers domains of both mirror types, in the first approximation this should result in an image of two overlapping reciprocal lattices  [Fig.~\ref{fig:LEED}(e)]. In this case spots at $\times2$ position will form twice denser rows than what is expected for a single domain. Taking into account the elongation of spots observed in the experiment and the expected higher intensity for spots that fall on top of each other [Fig.~\ref{fig:LEED}(f)], spots at $\times2$ position appear almost like lines with reduced intensity. This simplistic model shows a certain agreement with the pattern observed in the experiments, however, the lines in LEED do not look just like a series of short streaks, therefore additional arguments are required.
	
	Existence of two different wire alignments leads to breaking of the translational symmetry and and the long-range order in the direction perpendicular to the wire. This effect may lead to broadening of the LEED spots perpendicular to the wires. However, a more detailed analysis is needed to explain why only spots with \textit{odd} first indices become so broad that they appear as lines, while spots with \textit{even} first indices are just slightly broadened.
	
	In a more elaborate model one has to take into account that there are no large domains of one mirror type or another, instead the chains are stacked in one way or another nearly at random, since there is no visible preference of one structure over the other (it can be seen in the STM data of Ref.~\cite{BlueP-Zhang-NatComm-2021}). Such situation is well known in X-ray diffraction and the intensity can be calculated analytically (see e.g. Ref. \cite{Wilson1962}, Chapter V). The high probability of stacking faults with a shift of 1/2 of a unit cell results in a strong broadening of the diffraction peaks in the direction perpendicular to this shift \textit{but only for odd-order reflections}. And in the case of exactly 50\% probability of a stacking fault (i.e. one or another alignment are chosen randomly), the intensity distribution function becomes a constant, leading to uniform lines of intensity.
	Such mechanism was also suggested to explain similar effects of lines in LEED for antiphase domains in MoO$_2$ \cite{Schroeder2002} and dimerized Au chains on Si(553) \cite{Hafke2016}.
	
	\section{Details of the experiment, analysis and calculations}
	
	XPS and LEED measurements presented in the paper were performed at PEARL beamline (X03DA) of the Swiss Light Source.
	
	ARPES measurements were performed at the beamline UE112-PGM2 of BESSY II synchrotron using endstation ARPES 1$^2$ equipped with cryomanipulator S6.Cryo \cite{ARPES-robotics}. 
	
	Characterization by scanning tunneling microscopy (STM) was conducted with Omicron LT STM at $T=4.5$ K
	using tungsten tips prepared as described elsewhere \cite{Varykhalov-PRB-2005}. 
	Scanning tunneling spectroscopy (STS) measurements were executed using the lock-in technique. 
	
	Synthesis of P chains was achieved by controlled {\it in vacuo} sublimation of a BlackP crystal onto clean Ag(111) held at room temperature. 
	
	DFT calculations, including preliminary structural optimization, were performed using the 
	VASP package \cite{Kresse} with the PAW method in the PBE approximation and DFT-D2 
	van der Waals interaction correction \cite{Grimme}. Band unfolding was performed 
	using the {\it VaspBandUnfolding} code \cite{Unfolding}. The Ag(111) substrate
	was simulated as 12 atomic layers of silver.

	Parabolic fitting to determine the dispersion of the Ag(111) surface state between P chains was performed under the assumption that the parabola minimum is at $k=0$. Errors in determination of the wavelengths were taken into account using the trust-region Levenberg-Marquardt least orthogonal distance method as implemented in Igor Pro software utility.
	
	Atomic structures were modelled using the VESTA software utility \cite{VESTA}.
	
	STM and STS data were analyzed using the software utility Gwyddion \cite{Necas2012,Gwyddion}.
	
	LEED patterns in Fig. S3 were simulated using the software utility LEEDpat 4.2 \cite{LEEDpat42}.
	
	\bibliography{SI_P1DAg111}